\documentclass[iop,twocolumn,numberedappendix]{emulateapj}
\usepackage{graphicx}
\usepackage{url}
\usepackage{multirow}
\usepackage{amsmath}
\usepackage{enumerate}

\slugcomment{} 
\shorttitle{A quasar - galaxy cross}
\shortauthors{}
\begin{document}
\title{The quasar-galaxy cross SDSS~J1320+1644: A probable large-separation lensed quasar\altaffilmark{1}}
%
\author{
Cristian E. Rusu,\altaffilmark{2,3} 
Masamune Oguri,\altaffilmark{4,5} 
Masanori Iye,\altaffilmark{2,3,6} 
Naohisa Inada,\altaffilmark{7} 
Issha Kayo\altaffilmark{8} 
Min-Su Shin,\altaffilmark{9} 
Dominique Sluse\altaffilmark{10}
and 
Michael A. Strauss\altaffilmark{11} 
}

\altaffiltext{1}{Based on data collected at Subaru Telescope, which is
operated by the National Astronomical Observatory of Japan. Use of
the UH2.2 m telescope for the observations is supported by NAOJ.} 
\altaffiltext{2}{Optical and Infrared Astronomy Division, National
Astronomical Observatory of Japan, 2-21-1,
Osawa, Mitaka, Tokyo 181-8588, Japan.}  
\altaffiltext{3}{Department of Astronomy, Graduate School of Science,
University of Tokyo 7-3-1, Hongo Bunkyo-ku, Tokyo 113-0033, Japan} 
\altaffiltext{4}{Kavli Institute for the Physics and Mathematics of the
     Universe, The University of Tokyo, 5-1-5 Kashiwa-no-ha, Kashiwa, 
     Chiba 277-8568, Japan.}
\altaffiltext{5}{Division of Theoretical Astronomy, National
                Astronomical Observatory of Japan, 2-21-1, Osawa,
                Mitaka, Tokyo 181-8588, Japan.} 
\altaffiltext{6}{Department of Astronomical Science, The Graduate University
                for Advanced Studies (SOKENDAI), National Astronomical Observatory
                of Japan, 2-21-1,Osawa, Mitaka, Tokyo 181-8588, Japan} 
\altaffiltext{7}{Department of Physics, Nara 
	      National College of Technology, Yamatokohriyama,
	      Nara 639-1080, Japan}    
\altaffiltext{8}{Department of Physics, Toho University, Funabashi, Chiba
	       274-8510, Japan.}
\altaffiltext{9}{Department of Astronomy, University of Michigan, 500 Church Street, Ann Arbor, MI 48109-1042 USA.}
\altaffiltext{10}{Argelander-Institut f\"ur Astronomie, Auf dem H\"ugel 71, 53121 Bonn, Germany.}
\altaffiltext{11}{Princeton University Observatory, Peyton Hall, Princeton, NJ 08544, USA.}

\begin{abstract}
We report the discovery of a pair of quasars at $z=1.487$, with a separation of $8\farcs585\pm0\farcs002$. 
Subaru Telescope infrared imaging reveals the presence of an elliptical and a disk-like galaxy located almost symmetrically between the quasars, in a cross-like configuration. Based on absorption lines in the quasar spectra and the colors of the galaxies, we estimate that both galaxies are located at redshift $z=0.899$. This, as well as the similarity of the quasar spectra, suggests that the system is a single quasar multiply imaged by a galaxy group or cluster acting as a gravitational lens, although the possibility of a binary quasar cannot be fully excluded. We show that the gravitational lensing hypothesis implies that these galaxies are not isolated, but must be embedded in a dark matter halo of virial mass  $\sim 4 \times 10^{14}\ h_{70}^{-1}\ \mbox{M}_\odot$ assuming an NFW model with a concentration parameter of $c_{vir}=6$, or a singular isothermal sphere profile with a velocity dispersion of $\sim 670$ km s$^{-1}$. We place constraints on the location of the dark matter halo, as well as the velocity dispersions of the galaxies. In addition, we discuss the influence of differential reddening, microlensing and intrinsic variability on the quasar spectra and broadband photometry.
\end{abstract}


\section{Introduction}\label{sec:intro}

Quasar pairs belong to one of three categories: gravitational lenses, binary quasars, and apparent (projected) pairs. Both components of a pair in the first two categories have the same redshift\footnote{Small redshift differences in the components of binary quasar pairs, such as $661\pm173$ km s$^{-1}$ for LBQS 0015+0239 \citep{impey02}, are consistent with the line-of-sight velocity difference of bound pairs of galaxies.}. Gravitationally lensed quasars dominate the small separation $\Delta\theta\lesssim 3''$ quasar pairs. These are produced by galaxy scale lenses, and for the past three decades have proven to be invaluable astrophysical and cosmological probes \citep[e.g.,][]{claeskens02,schneider06}. At larger separations, the lensing probability falls quickly \citep{turner84}, and becomes increasingly dominated by environmental effects due to galaxy groups and clusters \citep[e.g.,][]{keeton00}. The largest separation, double-image lensed quasar confirmed to date is Q0957+561 \citep[$\Delta\theta=6\farcs17,$][]{walsh79}, and the only larger separation lensed quasars known are the five-image SDSS~J1004+4112 \citep[$\Delta\theta=14\farcs7,$][]{inada03,inada05} and the three-image SDSS~J1029+2623 \citep[$\Delta\theta=22\farcs5,$][]{inada06,oguri08-2}. The two latter systems are lensed by clusters. Clearly, a larger sample is needed to study observationally the lensing probability distribution in the large-separation range \citep[e.g.,][]{oguri06-2,more12}. These wide lenses provide unique opportunities to study the interplay between baryonic and dark matter in groups ($\Delta\theta\lesssim10''$), as well as the dark matter distribution in clusters\footnote{It is worth mentioning that, in addition to lensed quasars, there are also many known lensed galaxies, with a large range of image separations, which we do not address here \citep[e.g.,][]{faure08,auger09,more12}.} \citep[e.g.,][]{oguri04}.

Most large-separation quasar pairs known, however, are binary quasars \citep[e.g., 19 pairs with $3''<\Delta\theta<10''$ in the sample of][]{hennawi06}. These are physically associated quasars, in the sense that they are either gravitationally interacting, or belong to the same group or cluster. Binary quasars are found to be more abundant than predicted from extrapolating the quasar correlation function to small scales \citep[e.g.,][]{djorgovski91,hewett98,hennawi06,kayo12}, which may imply that quasar activity is triggered and sustained by tidal interactions or mergers \citep[e.g.,][and references therein]{mortlock99,hopkins05,hopkins06,hopkins07,hopkins08,green10}.

Many quasar pairs found in the range $3''\lesssim\Delta\theta\lesssim10''$ have proven difficult to classify as either lensed or binary quasars \citep[e.g. Q2345+007;][]{weedman82}. In these systems no plausible lenses have been identified in the foreground, although the redshifts of the two components are identical and the spectra are quite similar. The spectra of gravitationally lensed images need not necessarily be identical, since they are prone to effects such as delayed intrinsic variability, chromatic microlensing, and differential extinction in the lensing galaxy \citep[e.g.,][]{wucknitz03,yonehara08}. On the other hand, there is little dynamic range in the spectral variation of quasars, meaning that pairs can have similar spectra by chance \citep[e.g.,][]{mortlock99}. 

The suggestion of lensing by "dark" lenses \citep[e.g.,][]{koopmans00} or cosmic strings \citep{vilenkin84} has been put forward for these quasar pairs. There are in fact very few criteria for discriminating between the lensing and binary quasar hypotheses, without monitoring the luminosity variability of the components, in order to look for a time delay. Instead, clues about this population of pairs have been found statistically. For instance, it has been argued that most of these quasar pairs are not lensed quasars, based on comparisons of the optical and radio properties of the population \citep[][]{kochanek99}. As for individual pairs, their nature must in general be estimated from the ensemble of their observed properties.

In this paper, we report the discovery in the course of the Sloan Digital Sky Survey Quasar Lens Search \citep[SQLS;][]{oguri06-1,oguri08-1,oguri12-2,inada08,inada10,inada12} of a new large-separation ($\Delta\theta\sim8\farcs5$) quasar pair which has proven difficult to classify, SDSS~J132059.17+164402.59 and SDSS~J132059.73+164405.6 (hereafter jointly SDSS~J1320+1644). In the near infrared, we detect two galaxies located between the quasars, producing an apparent cross-like configuration.  
In \S\ref{sec:sdss}, we give a brief description of the lens candidate selection from the SDSS data, and in  \S\ref{sec:obs} we present our imaging and spectroscopic follow-up observations. Next, we estimate photometric redshifts for the two galaxies and show that they are located in the foreground (\S\ref{sec:photoz}). We subsequently proceed with the gravitational lens mass modeling of this system (\S\ref{sec:mass}). In \S\ref{sec:discuss-spec}, we examine the quasar spectral and photometric differences, whereas in \S\ref{sec:discuss-cluster} we explore the environment of the system. Finally, we summarize our conclusions in \S\ref{sec:summary}. 
Throughout this paper, we assume the concordance cosmology with $H_0=70\ h_{70} \ \mathrm{km}^{-1}\
\mathrm{s}^{-1}\ \mathrm{Mpc}^{-1}$, $\Omega_M=0.27$ and $\Omega_\Lambda=0.73$. 

\section{Discovery in the Sloan Digital Sky Survey}\label{sec:sdss}

The Sloan Digital Sky Survey \citep[SDSS-I, 2000-2005, SDSS-II, 2005-2008;][]{york00} is a combination of imaging  and spectroscopic surveys which have mapped $\sim$ 10,000 square
degrees of the sky, centered at the North Galactic Cap.  
The observations are made at the Apache Point Observatory in New Mexico, USA, with a dedicated 2.5-meter wide-field telescope \citep{gunn06}.   
The imaging survey is conducted in five broad-band filters, centered at 3561($u$), 4676 ($g$), 6176 ($r$), 7494 ($i$), and 8873{\,\AA} ($z$) \citep{fukugita96,gunn98,doi10}.
The automated pipeline reduction achieves an astrometric accuracy better than about
$0\farcs1$ \citep{pier03}, and a photometric zeropoint accuracy better
than about 0.01 magnitude over the entire survey area, in the $g$, $r$, and $i$ bands 
\citep{hogg01,smith02,ivezic04a,tucker06,padmanabhan08}. 
The spectroscopic observations are carried out with a multi-fiber
spectrograph between $3800-9200${\,\AA}, with a resolution of
$R\sim1800-2100$ \citep{blanton03}. All data have been made publicly available, in periodic data releases \citep{stoughton02,abazajian03,abazajian04,abazajian05,abazajian09,
adelman06,adelman07,adelman08}. 

The SQLS has been one of the most successful strong lens surveys conducted to date. It identifies lensed quasar candidates among spectroscopically confirmed quasars in the SDSS \citep{richards02,schneider10}, by combining two selection algorithms, designed to find both small- and large-separation lenses, respectively. The morphological selection identifies small-separation candidates that are unresolved, yet poorly fitted by the point spread function (PSF); the color selection algorithm searches, in the vicinity of every confirmed quasar, for resolved stellar objects that have similar colors to the quasar. The SQLS has discovered more than 40 new lensed quasars so far  \citep[][]{inada12}. Together with previously known lenses rediscovered by the SQLS, it has produced a sample of $\sim 60$ lensed quasars, roughly half of all lensed quasars discovered to date. 

SDSS~J1320+1644 is a large-separation lensed quasar candidate identified by the color selection algorithm. We reproduce the SDSS multi-band composite image of the system in Figure \ref{fig:sdss-imag}. A faint red object, classified by the automatic pipeline as a galaxy, is detected in between the two blue stellar components constituting the lensed quasar candidate, towards North. We give the SDSS photometric results for the quasars and the red object in Table \ref{tab:sdss-phot}. As a result of the finite size of the optical fibers in the SDSS spectrograph, spectra are usually not obtained for two objects less than $55''$ apart \citep[the "fiber collision";][]{blanton03}, and therefore SDSS spectroscopy is available for only one member of the pair. We reproduce the spectrum of that object, classified as a quasar at $z=1.5024\pm0.0024$, in Figure \ref{fig:spec} (in blue).

\section{Follow-up Spectroscopy and Imaging}\label{sec:obs}

\subsection{Spectroscopy of the quasar pair}\label{sec:spec}

Follow-up spectroscopy aimed to ensure that both stellar components are quasars at the same redshift, and to assess their spectral similarities, was performed at the Astrophysical Research Consortium Telescope (ARC 3.5 m), located at the Apache Point Observatory. Spectra were taken with the Dual Imaging Spectrograph, equipped with a combination of blue and red gratings, B400/R300. 
The observations were conducted on 2009 February 19 with a $1\farcs5$ width slit oriented so that both stellar components are on the slit. One 1500 s exposure was taken at an effective airmass of 1.63. The extracted spectral coverage was $3700 - 10000$ \AA, with a resolution R $\sim 500$. The data reduction was performed using standard IRAF\footnote{The interactive Reduction and Analysis Facility (IRAF) is distributed by the National Optical Astronomy Observatories, which are operated by the Association of Universities for Research in Astronomy, Inc., under cooperative agreement with the National Science Foundation.} tasks.

The spectra shown in Figure \ref{fig:spec} (black and red) indicate that both stellar components have similarly shaped quasar broad emission lines (BELs; \ion{C}{3]}, \ion{Mg}{2}) at the same wavelengths, and therefore are quasars at the same redshift. Our redshift estimate based on the \ion{C}{3 ]} and \ion{Mg}{2} peaks (including a reanalysis of the original SDSS spectrum, which was taken in a longer, 4200s exposure, and better resolution R $\sim 1850-2000$) is 1.487$\pm0.001$, which is slightly lower than, and incompatible with the value in the SDSS database and its associated error bar. We identify absorption lines in both quasars as the \ion{Mg}{2} 2796 \AA, 2803 \AA\ doublet and possibly \ion{Fe}{2} 2599.4 \AA,  indicating $z=0.899$. We identify another \ion{Mg}{2} doublet at $z=1.168$, present only in quasar B (the quasar designation corresponds to that in Figure \ref{fig:subaru-imag}).

The flux ratio A/B increases with wavelength, from $\sim$ 0.8 to $\sim$ 1.5. This could be explained by a combination of microlensing, differential extinction and intrinsic variability, or simply differential slit losses, and does not constitute by itself an argument against the gravitational lensing hypothesis. We provide a more thorough comparative analysis of the spectra in \S\ref{sec:discuss-spec}.

Overall, the identical redshift of the two quasars, the similarity of the shapes of the BELs, as well as the presence of absorption lines at the same redshift, appear consistent with the gravitational lensing hypothesis. Of course, these features are fully consistent with the binary quasar hypothesis as well, in which case spectral differences arise naturally, as there are two distinct quasars.

\subsection{UH88 and Subaru Telescope imaging data}\label{sec:image}

Optical follow-up imaging observations were originally conducted with the Tektronix 2048$\times$2048 CCD camera (Tek2k; $V$, $R$, $I$ and $z$ bands) and the Wide Field Grism Spectrograph 2 (WFGS2; $z$ band) at the University of Hawaii 2.2-meter (UH88) telescope, in April 2009. The Tek2k camera has a $7'.5\times7'.5$ field of view and a pixel scale of $0\farcs219$~pixel$^{-1}$. The WFGS2 has a $11'\times11'$ field of view and focal reducer of $0\farcs34$~pixel$^{-1}$, and was used in imaging mode with the Tek2k camera. Bias and flat frames were also obtained, and the data were reduced using standard IRAF tasks. The standard star system PG0918+029 \citep{landolt92} was observed for photometric calibration, but the zero-points determined from the five individual stars have a scatter of $\sim 0.2$ mag, possibly indicative of non-photometric conditions. In order to obtain a better precision, the instrumental magnitudes of 10 bright stars in the field of view were compared to their SDSS \emph{ugriz} magnitudes. The Lupton (2005) formulae\footnote{\protect\url{http://www.sdss.org/dr5/algorithms/sdssUBVRITransform.html}}  were used to transform between \emph{ugriz} and $VRI$ magnitudes, which resulted in smaller scatter. The exposure time, airmass, seeing, zero-point uncertainty and observation date are given for each band in Table \ref{tab:followup-data}. 

\begin{figure}
\epsscale{0.5}
\plotone{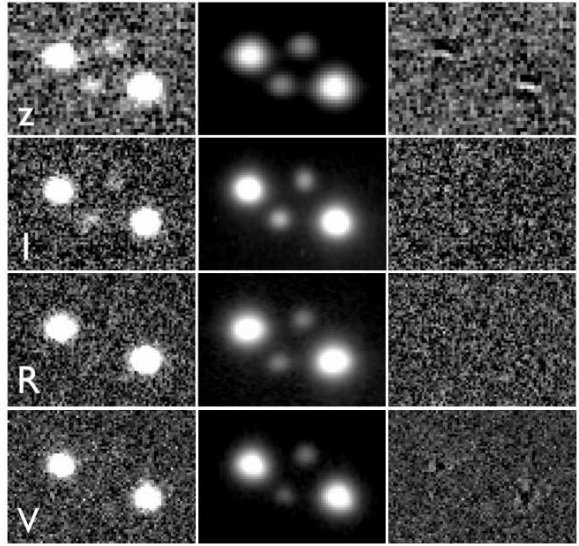}
\caption{Imaging of SDSS~J1320+1644 with the UH88 telescope. In each of the $V$, $R$, $I$ and $z$ bands, the images to the left show the original observations, the ones in the center show the best models obtained with GALFIT, and those to the right show the residuals after subtracting the fitted models. The snapshots are $11\farcs8\times16\farcs6$. North is up and East is to the left.
\label{fig:VRIz}}
\end{figure}

In the deeper UH88 images (the $I$ and $z$ filters), in addition to the faint photometric object detected between the two quasars in the SDSS data, another object is detected, located almost symmetrically with respect to the quasars (Figure \ref{fig:VRIz}). In order to determine the morphological and photometric properties of these two objects, as well as to estimate photometric redshifts, we obtained higher resolution near-infrared ($J$, $H$ and $Ks$ bands) images with the Multi-Object InfraRed Camera and Spectrograph \citep[MOIRCS;][]{ichikawa06,suzuki08}, at the Subaru 8.2-meter telescope \citep{iye04}, in April 2010. MOIRCS has two HAWAII-2 2048$\times$2048 detectors providing a  $3'.94\times6'.90$ total field of view, with a pixel scale of $0\farcs117$~pixel$^{-1}$. The data (Table \ref{tab:followup-data}) were reduced with the MCSRED software package (I. Tanaka et al., in preparation), and the standard star FS33 \citep{leggett06} was used to estimate the photometric zero points.

In the $J$, $H$ and $Ks$ imaging (Figure \ref{fig:subaru-imag}), the two faint objects are clearly detected. They have extended morphology and red colors, indicative of fairly high-redshift galaxies. In each band, we modeled all four objects simultaneously using the public software GALFIT
\citep{peng02}, with nearby bright stars as PSF templates. We fitted the galaxies with a S\'{e}rsic profile convolved with the PSF. The resulting astrometry and photometry (including the UH88 photometry) are given in Tables \ref{tab:astrometry} and \ref{tab:photometry}, respectively. We were able to characterize the morphology of the two galaxies, which we summarize in Table \ref{tab:morphology}. The S\'{e}rsic index of G1 is close to the canonical value for elliptical galaxies, $n=4$, whereas for G2 it is close to the typical value for disk galaxies, $n=1$. We therefore assume that G1 is an elliptical galaxy, whereas G2 is disk-like. This is also consistent with the bluer color of G2. 

We also detected three additional fainter objects in the vicinity of the system. While FWHM measurements and GALFIT modeling are not very reliable, our analysis suggests that the objects are extended, and so we mark them as galaxies G3, G4 and G5 in Figure \ref{fig:subaru-imag}. Aperture photometry, summarized in Table \ref{tab:photometry}, indicates that the colors of G4 (but not of G3 and G5) are very similar to those of G1 and G2, and we therefore assume that this is a galaxy at about the same redshift. We also remark that the $J-H$ colors of G3 and G5 are different from those of A and B, and it is therefore unlikely that these are additional, faint gravitationally lensed quasar images.

In the UH88 and Subaru Telescope imaging observations, the flux ratio A/B is above unity, in opposition to the SDSS results. The chromatic change in the brightness ranking of the quasar images could be explained by microlensing (see \S\ref{sec:microlensing}).

\section{Photometric redshifts}\label{sec:photoz}

In order to determine whether the two brighter galaxies are located in the foreground of the quasar pair, as is required by the gravitational lens hypothesis, we need to estimate their redshifts. We estimate photometric redshifts, based on the magnitudes of the galaxies in different filters, using the template-fitting methods implemented in the publicly available HyperZ \citep{bolzonella03} and EAzY \citep{brammer08} algorithms. These algorithms fit the observed magnitudes to spectral energy distribution (SED) templates, via $\chi^2$ minimization. As templates, we employ the observed mean SEDs of local galaxies from \citet[][hereafter CWW]{coleman80}, which are extrapolated into the ultraviolet and near-infrared with the evolutionary models of \citet{bruzual93}. The EAzY algorithm also uses the redshift distribution of galaxies of a given apparent magnitude as a Bayesian luminosity prior, following \citet{benitez00}. This helps to break the degeneracies between multiple probability distribution peaks at different redshifts. 

In order to obtain consistent results for the two algorithms, we had to increase the uncertainties on the magnitudes to about double the values quoted in Table \ref{tab:photometry}, which also makes most of the aperture- and model-determined values consistent. In addition to the statistical errors quoted in that table, additional sources of errors could potentially be introduced by the zero-point uncertainties (Table \ref{tab:followup-data}), possible blending of sources due to their close proximity and large seeing, different pixel size in the $V$, $R$, $I$ and $z$ band observations (aperture photometry), or the modeling difficulties and PSF uncertainties (model magnitudes). 

\begin{figure}
\epsscale{1}
\plotone{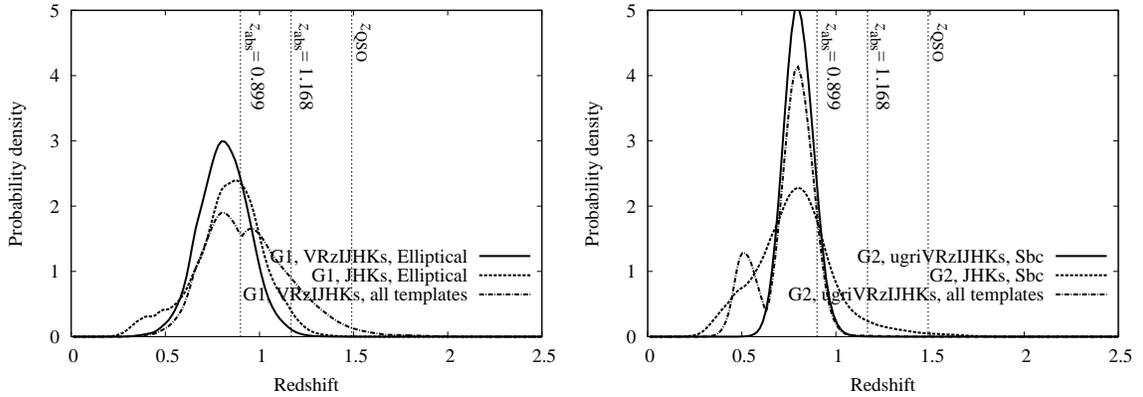}
\caption{Photometric redshift probability distributions, normalized to unit area,
for the two galaxies G1 ({\it left}) and G2 ({\it right}). The probability distributions are calculated with EAzY using the magnitudes in Tables \ref{tab:sdss-phot} and \ref{tab:photometry} (see also \S\ref{sec:photoz} for details) for the elliptical template, Sbc template, or marginalized over all spectral templates. The redshift of the quasar pair, as well as those of the absorption line systems identified in the quasar spectra, are marked with vertical lines. 
\label{fig:photoz}}
\end{figure}

The results of the photometric redshift estimates are shown in Figure \ref{fig:photoz} and Table \ref{tab:photoz}. G1 is well-fitted by an elliptical template, and G2 by an Sbc template, consistent with the expectations from the galaxy morphologies (see \S\ref{sec:image}). Although the probability distributions are rather broad, they are in agreement, inside the $1\sigma$ interval estimated from the redshift probability distributions, with $z=0.899$, which is the redshift of the absorption lines identified in both spectra. The redshift of the second \ion{Mg}{2} absorption system in quasar B, at $z=1.166$, is rejected at about $2\sigma$ confidence level, while the probability of a redshift as high as the quasar redshift appears negligible. Therefore, throughout the rest of this work, we assume that these galaxies are responsible for the absorption lines at $z=0.899$.

For the faintest objects we detected, G3, G4 and G5, the redshift probability distributions are very broad (Table \ref{tab:photoz}). G3 and G5 show a smaller preferred redshift of $z \sim 0.5$. On the other hand, G4 has a preferred redshift of 1.11, very close to that of the \ion{Mg}{2} absorption system at $z=1.168$. This absorption system is quite strong, with equivalent width $\sim 2.9$\ \AA\  in the rest frame of the absorber. Here the projected proper distance between B and G4 is 21$\ h_{70}^{-1}$  kpc at $z=1.168$. On the other hand, the \ion{Mg}{2} absorption system at $z=0.899$ has equivalent widths $\sim 2.9$\ \AA\ and $\sim 4.5$\ \AA\ (in the spectra of quasars A and B, respectively) for impact parameter $\sim 35\ h_{70}^{-1}$ kpc, corresponding to the distances between the quasars and G1, G2. Such strong absorptions at small impact parameters are consistent with the literature \citep[e.g., ][and references therein]{churchill05}. The stronger absorption in quasar B could be explained by its closer proximity to one of the galaxies (G1).

\section{Gravitational Lens Mass Modeling}\label{sec:mass}

We next proceed with modeling the system as a gravitational lens. The observational constraints provided by the available astrometry and photometry are the positions of the images A and B, their fluxes (we take the flux ratio to be either $\sim1.0 \pm 0.2$ or $\sim1.4 \pm 0.2$, as we will show in \S\ref{sec:discuss-spec}), and the positions of the two main galaxies expected to act as lenses, G1 and G2. We also know the redshift of the source, $z=1.487$, and of the lenses, $z=0.899$. As the number of observational constraints is limited, we focus on simple SIS and NFW lens models and their elliptical counterparts, with or without shear. The singular isothermal profile is known to be a good approximation for the mass distribution in galaxies \citep[e.g., ][and references therein]{rusin05}. It has also been employed in the lens modeling of galaxy clusters \citep[e.g., ][]{inada03}. The NFW profile \citep[][]{navarro96}, on the other hand, is frequently employed as an approximation of the mass distribution in dark matter haloes, which dominate galaxy clusters. Constraints can also be placed on the "strength" of the lenses, represented (in the case of the SIS model) by their velocity dispersions, which we infer from the observed luminosities. To this end, we estimate the rest frame $R$ band absolute magnitudes of G1 and G2 as $\sim -22.5$ and $\sim -22.3$, respectively, calculated with HyperZ from the best-fit E and Sbc templates. We use the Faber-Jackson \citep[F-J;][]{faber76} and the Tully-Fisher \citep[T-F;][]{tully77} laws, with parameters given in \citet[][]{binney08}, and infer velocity dispersions of $237\pm54$ km s$^{-1}$ and $163\pm15$ km s$^{-1}$, respectively. For G2, which is disk-like, we have assumed a singular isothermal sphere (SIS) profile and estimated the velocity dispersion as $\sigma=v_{\rm circ}/\sqrt{2}$, where $v_{\rm circ}$ is the circular velocity obtained from the T-F law. 

The velocity dispersion estimate above implies that the brightest galaxy G1 has an Einstein radius of only $\sim 0\farcs5 \pm 0\farcs$2, and an inclosed mass of $\sim (2.1 \pm 1.5) \times10^{11}\ h_{70}^{-1}$ ${\rm M}_\odot$. However, the large separation between A and B ($\sim8\farcs6$), assuming they are lensed images of a single source, implies an enclosed mass of $\sim 1.2\times10^{13}\ h_{70}^{-1}$ ${\rm M}_\odot$. Consequently, this cannot be a gravitational lens system where the main deflectors are the individual members of a group of galaxies, and we need to invoke a significant embedding dark matter halo.

A question worth posing is to what extent must the faintest objects G3, G4 and G5, be taken into account when building gravitational lens models. We only estimate the velocity dispersion of G4, assumed to be at $z=0.899$ or $z=1.168$ (see \S\ref{sec:photoz}). We neglect the contribution of G3 and G5, as they are located further away from images A and B, and are most likely at a significantly different redshift from G1, G2 and G4. Assuming that G4 is located at the same redshift as G1 and G2 leads to velocity dispersions $\sigma=150\pm35$ km s$^{-1}$ or $\sigma=105\pm10$ km s$^{-1}$, from the F-J and T-F laws, respectively. A smaller effective velocity dispersion would be inferred if placing G4 at redshift $z=1.168$ (see \S\ref{sec:appen-G4} for details). 

We subsequently explore gravitational lens models in which the lensing potentials of G1, G2 and G4 are boosted by an embedding dark matter halo. We model the observed galaxies with SIS profiles, and the dark matter halo with either an SIS or an NFW profile. We consider the case where the dark matter halo is centered at the location of the brightest galaxy G1, as well as the case where its position is a free parameter, on which we obtain constraints. We only consider models with 0 degrees of freedom (d.o.f.), which we are able to fit perfectly ($\chi^2\ll1$). This means that for the models where the center of the dark matter halo is fixed, we save two parameters which we can use to introduce a shear/ellipticity and associated position angle. The gravitational lens modeling is performed with {\it glafic} \citep{oguri10}. 

\begin{figure}
\epsscale{1}
\plotone{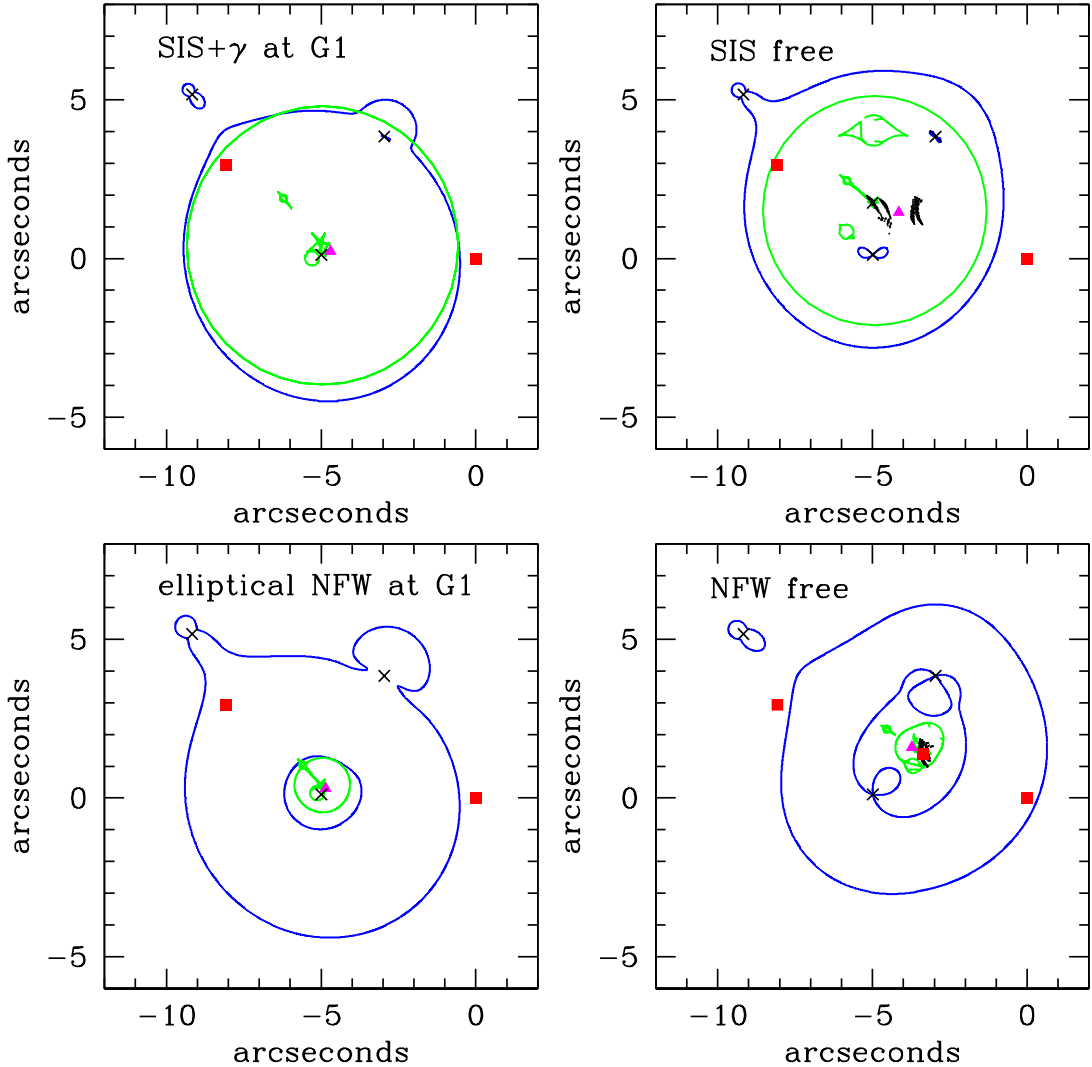}
\caption{Gravitational lensing models. {\it Top left}: A and B are two images of a source lensed primarily by a large velocity dispersion SIS (with shear) profile centered on G1;  {\it top right}: same as previous, but the position of the primary SIS (without shear) lens is not fixed; {\it bottom left}: A and B are two images of a source lensed primarily by a large elliptical NFW profile centered on G1; {\it bottom right}: same as previous, but the position of the spherical NFW lens is not fixed. Critical lines and caustics are drawn in blue (outer curves) and light green (inner curves), respectively, with the velocity dispersions of G1, G2 and G4 fixed at their most probable values (237, 163 and 118 km s$^{-1}$, respectively; see \S\ref{sec:mass} and \S\ref{sec:appen-cvir}), unless the adopted model requires otherwise. Only NFW profiles with $c_{vir}=6$ are drawn. Symbols denote image, source and lens positions (red squares, magenta triangles and black crosses, respectively). The black regions close to the center of the large caustic of the free position models show the locations of this primary lens, as the velocity dispersions of G1, G2 and G4 are varied over the allowed $1\sigma$ range ($\chi^2\leq2.3$). [\it{See the electronic edition of the Journal for a color version of this figure.}]
\label{fig:lens1}}
\end{figure}

We show the critical lines and caustics for these models in Figure \ref{fig:lens1}. We report the best fitted parameters of the models (velocity dispersion or virial mass of the dark matter halo, ellipticity/shear and associated position angle, total magnification and time delay) in Table \ref{tab:models}. The velocity dispersions of G2 and G4 (also G1 in the case of the "SIS free" and "NFW free" models of Figure \ref{fig:lens1}) were fixed at the values inferred earlier, but we also explored, on a suitable grid, combinations of velocity dispersions in the ranges allowed by the F-J and T-F laws. In this case, we only considered models with $\chi^2\leq2.3$, corresponding to the $1\sigma$ confidence interval for two parameters. The effect of the range of velocity dispersions is reflected in the range of values for the parameters in Table \ref{tab:models}, as well as the allowed position of the dark matter halo, for the "SIS free" and "NFW free" models.

There are two major differences between the NFW and SIS models considered for the dark matter halo. First, we parameterize the NFW profile by its virial mass M$_{vir}$ and concentration parameter $c_{vir}$ (as opposed to just a single parameter, $\sigma$, for the SIS model), and we fix $c_{vir}$ at a fiducial value 6, close to the typical value  for galaxy clusters at the mass range and redshift we are interested in \citep[e.g., ][and reference therein]{binney08}. We explore different choices of $c_{vir}$ in \S\ref{sec:appen-cvir}. Second, the NFW profile is shallower than SIS at the center, and can therefore produce an additional faint central image. Therefore, for the "NFW free" model, we require that the model produce no more than three images, and that the central image be $\sim50$ times or more fainter than B, corresponding to our detection limit at S/N $\sim5$ in the deepest band, $Ks$. The central image cannot be one of the faint objects G3 or G5, as those objects are too far away from the midpoint between A and B, are also most likely extended, and have different colors.

Finally, in addition to the models where the dark matter halo is fixed on G1, we tried a similar range of SIS$+\gamma$ models at the location of G2, but were unable to produce good fits. We note that G2 is farther away from the line connecting images A and B, and therefore a model centered on G2 is expected to do a poorer job of fitting the data. We do not try to fix the dark matter halo at the location of G4, as we are not sure whether this galaxy is at the same redshift as G1 and G2, and we also do not expect the dark matter halo to be centered on a fainter galaxy of the group/cluster.

Our main result in this section is that any reasonable mass model reproduces the observed image configuration. In order for this system to be a two-image gravitationally lensed quasar, we do however require a massive dark matter halo in which the visible galaxies are embedded. This halo should be centered either on G1 (but not on G2) or at some location close to the midpoint between A and B. The choice of the intrinsic flux ratio (either $\sim1.0$ or $\sim1.4$) does not affect the successful mass models qualitatively. Finally, in \S\ref{sec:appen-caveat} we discuss several caveats regarding the estimated velocity dispersions of the galaxies.

\section{Quasar Spectro-Photometric Differences}\label{sec:discuss-spec}

The three physical phenomena known to modify the spectra of the images of gravitationally lensed quasars are differential extinction due to dust in the lensing galaxy \citep[e.g.,][]{falco99}, microlensing due to stars in the lens \citep[e.g.,][]{chang79,schneider06}, and intrinsic quasar variability coupled with the time delays between images. The first process is a static one, whereas the other two are time-dependent. We assess their effects on the continuum and broad-band fluxes in turn.

\subsection{Differential extinction}\label{sec:reddening}

\begin{figure}
\epsscale{.5}
\plotone{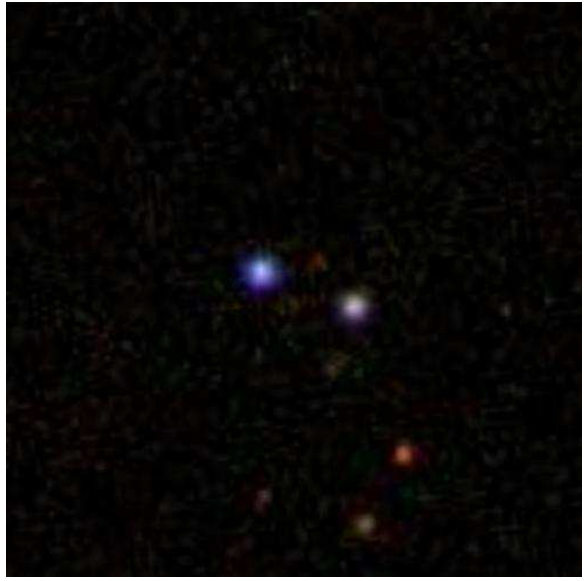}
\caption{SDSS $gri$ color-composite \citep{lupton04} centered on the SDSS~J1320+1644 pair, obtained on 2005, June 6. The image scale is $50''\times50''$. North is up and East is to the left. The exposure was 54 s in each band and the SDSS pixel scale is $0\farcs396$. The two quasars are the blue objects near the center. [\it{See the electronic edition of the Journal for a color version of this figure.}]
\label{fig:sdss-imag}}
\end{figure}

\begin{figure}
\epsscale{0.8}
\plotone{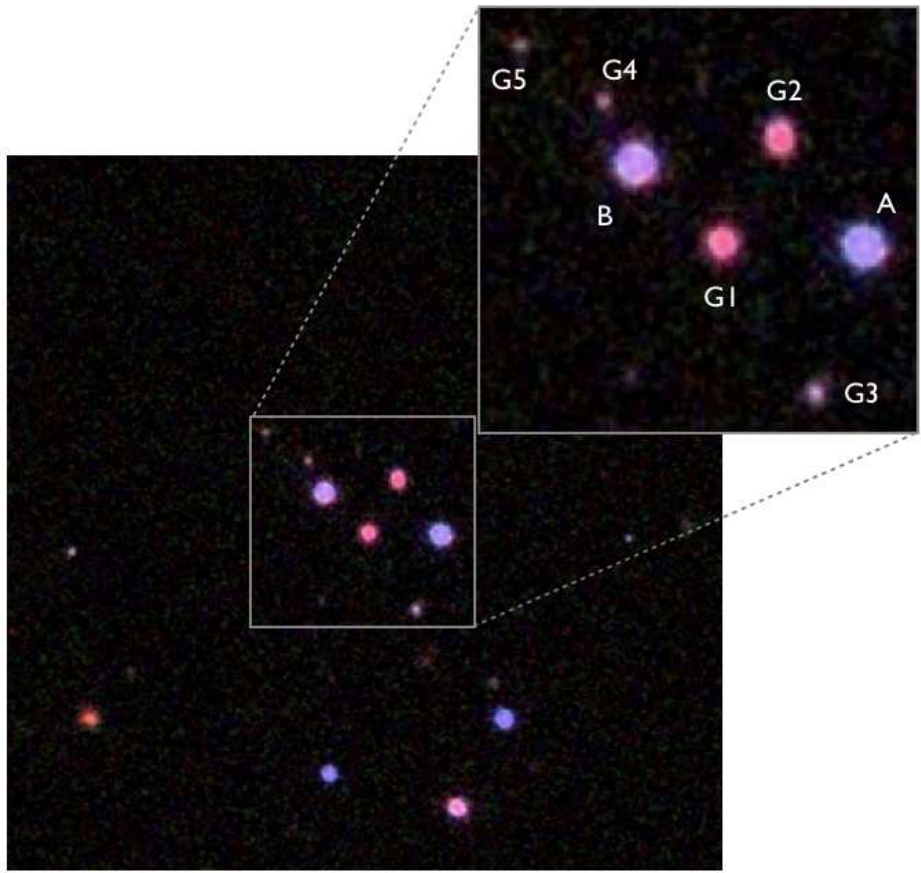}
\caption{Subaru Telescope imaging of SDSS~J1320+1644. The image scale (without the close-up) and orientation correspond to Figure \ref{fig:sdss-imag}. This is a small fraction of the MOIRCS field of view centered on SDSS~J1320+1644 (see the panel to the right in Figure \ref{fig:colorscluster}). The image is a color-composite of the $J$, $H$ and $Ks$ frames, using the algorithm of \citet{lupton04}. [\it{See the electronic edition of the Journal for a color version of this figure.}]
\label{fig:subaru-imag}}
\end{figure}

\begin{figure}
\epsscale{0.7}
\plotone{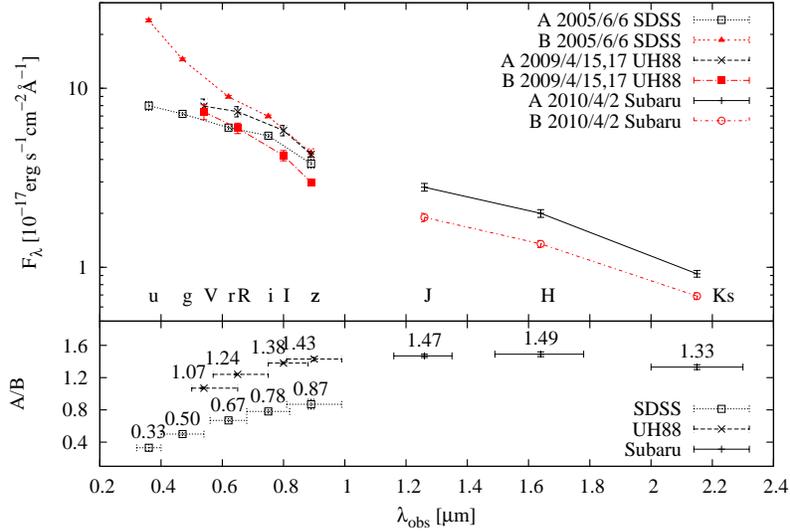}
\caption{Broadband photometric flux and flux ratios of quasars A and B, determined from the aperture photometry in Tables \ref{tab:sdss-phot} and \ref{tab:photometry}, which include curve of growth corrections. The photometric bands are marked, and the horizontal bars of the flux ratios indicate the width of the filters. We use the photometric errors corresponding to the zero-point uncertainties, from Table \ref{tab:followup-data}. The WISE mid-infrared data from Table \ref{tab:wise}, which is considered less reliable (see \S\ref{sec:variability}), is not plotted. [\it{See the electronic edition of the Journal for a color version of this figure.}]
\label{fig:flux}}   
\end{figure}

The most striking difference in the follow-up spectra of quasars A and B is the gradual drop in the flux ratio A/B with wavelength, shortward of about 7000 \AA. Such an effect has previously been observed in several lensed quasars, such as SDSS~J1001+5027 \citep[]{oguri05} and SDSS~J1313+5151 \citep[]{ofek07}. Differential extinction (reddening) between the quasar images could explain this effect, as it is most effective at short wavelengths. Here we caution however that the monotonous drop in the flux ratio may also be due to unknown differential slit losses in the ARC 3.5 m spectroscopy. Nonetheless, we do not expect the differential slit losses to be large, because the flux drop is supported by the available photometry from $u$ to $z$ band, which shows quasar A to be redder than B (see also Figures \ref{fig:sdss-imag} and \ref{fig:flux}). In addition, the flux ratios inferred from the UH88 photometry, obtained just two months after the follow-up spectra, are consistent with the spectral ratios at the respective wavelengths. On the other hand, extinction should be weakest in the longest wavelength filter $Ks$, which corresponds roughly to $J$ band in the rest frame of G1 and G2. A caveat however is that there is no obvious galaxy responsible for the extinction in A, i.e. located much closer to A than the other galaxies are to B.

In contrast, the large change in the spectral slope of quasar A at short wavelengths, on a time span of 13 months, between the original SDSS and ARC observations, requires a significant contribution of the time-dependent phenomena. While the slope change can also be caused by slit losses, the two sets of photometric data in the $V$ - $z$ band range, taken $\sim$ 4 years apart, also shows the brightness rank of the two quasars interchanging, with A getting brighter, and B becoming fainter. Therefore time-dependent phenomena must be acting on the quasars. We next look into the extent to which microlensing can explain these data. 

\subsection{Microlensing}\label{sec:microlensing}

We assume that the microlensing variability is due to the motion of stars in the lensing galaxies, which we consider to be dominated by the proper motion of the galaxies, with transverse velocity $v_\perp\sim\sigma$, where $\sigma$ is the velocity dispersion in the galaxy group/cluster. Following \citet[]{richards04} in the following formulae, there are two timescales to the microlensing event. The first is the expected duration of the event, which is the time in which the microlens crosses the source,

\begin{equation}
t_\mathrm{src}\sim4.1\left(\frac{R_\mathrm{src}}{1\ \text{lt-day}}\right)\left(\frac{v_\perp}{700\ \mathrm{km\  s^{-1}}}\right)^{-1} \mathrm{yr} , 
\label{eq:tsrc}
\end{equation}

\noindent where $R_\mathrm{src}$ is the size of the source quasar accretion disk in the rest frame ultraviolet, corresponding to the optical region of the observed spectrum of SDSS~J1320+1644. It has an estimated extent of a few light-days or less \citep[e.g.,][]{morgan08}.

The second timescale is the mean interval between microlensing events, which is crudely estimated as the time needed to cross an Einstein radius, 
\begin{equation}
t_\mathrm{Ein}\sim8.2\left(\frac{M}{0.1\ M_\odot}\right)^{1/2}\left(\frac{v_\perp}{700\ \mathrm{km\ s^{-1}}}\right)^{-1}\ h_{70}^{1/2}\ \mathrm{\ yr} , 
\label{eq:tein}
\end{equation}

\noindent with $M$ being the mass of the microlens.

Although these results are rough estimates, they are close to the timescales on which our data shows spectral and photometric variations. Careful analytical considerations of the expected microlensing light curves \citep[e.g.,][]{yonehara08} as well as ray-shooting techniques to account for the fact that the source could be microlensed by an ensemble of stars at any particular time \citep[e.g.,][]{kayser86} show that microlensing can indeed cause variations in amplitude comparable to, or even larger than we observe in the spectro-photometric data. In standard accretion disk theory \citep[e.g.,][]{peterson97}, short wavelength emission originates at comparatively smaller distances from the central engine, and so is preferentially affected by microlensing. This is in agreement with the large changes we detect at progressively shorter wavelengths in our two-epoch spectra of quasar A. Conversely, since the near-infrared emission is expected to originate at  larger distances from the center than the UV or optical continuum, we expect the longest wavelength band $Ks$ (originating approximately at $z$ band in the quasar rest frame) to be less affected by microlensing. 

On the other hand, the BEL region is as large as a few light-weeks or light-months \citep[e.g.,][]{clavel91}, with lower ionization lines originating at larger distances. Thus, we expect that, in particular, the \ion{Mg}{2} BEL is less prone to microlensing effects. We find that the spectral flux ratio A/B of the two images (Figure \ref{fig:spec}) shows a decrease at the center of the \ion{Mg}{2} BEL, compared to the surrounding continuum\footnote{This does not refer to the increase in A(SDSS)/B and A(SDSS)/A at the blue wing of the \ion{Mg}{2} BEL, which is due to telluric absorption calibrated out in the SDSS spectrum, but not in the follow-up spectra.}. This could be indicative of microlensing affecting the continuum, but not the \ion{Mg}{2} BEL.

In principle, an estimate of the intrinsic flux ratio can be obtained from the flux ratio in the continuum-subtracted BELs, since they are less prone to microlensing effects than the continuum fluxes. However, in our case there is the chance that differential slit losses, although believed to be small (see \S\ref{sec:reddening}) affect the BELs, and therefore we refrain from doing so. 

One possible argument against microlensing could be that, since the two galaxies are fairly distant in projection from the quasar pair ($\sim 35\ h_{70}^{-1}$ kpc), there may be no stars to account for microlensing. We note that evidence of microlensing has previously been found in both SDSS~J1004+4112 \citep[e.g.,][]{richards04}, and SDSS~J1029+2623 \citep[e.g.,][]{oguri08-2}, the two large-separation cluster lenses; in those cases however, the galaxies believed to host the stars responsible for microlensing are located at slightly smaller projected separations of $\lesssim 20 - 30\ h_{70}^{-1}$ kpc from the quasar images.

\subsection{Intrinsic variability}\label{sec:variability}

We estimate the magnitude of intrinsic quasar variability using the structure function \citep[e.g.,][and reference therein]{yonehara08}

\begin{equation}
V=(1+0.024M_i)\left(\frac{\Delta t_\text{RF}}{\lambda_\text{RF}}\right)^{0.3} \text{\ mag.}
\label{eq:intrinsic}
\end{equation}

\noindent This predicts the magnitude change for a quasar of $i$ band absolute magnitude $M_i$, at a given rest frame wavelength $\lambda_\text{RF}$ in units of \AA, for a rest frame time interval between observations of $\Delta t_\text{RF}$ (in days). The structure function has been measured from a quasar sample with $M_i \sim-21$ to $\sim-30$ \citep[][]{vandenberk04}, in the rest frame optical/UV, and on a rest frame time range of up to 2 years. We also note that the characteristic time scale for optical variability of quasars is of the order 1 year \citep[][]{ivezic04b}, which is comparable with the time delay in the source rest frame (0.4 - 2.8 years), in our best lensing models. 

We use the structure function to check if it can reproduce spectral changes in quasar A of a factor of $\sim2$ in flux ($\sim$ 0.75 mag), at $\sim4500$ \AA\ ($\sim1800$ \AA\ rest frame), in the 13 months ($\sim 160$ days rest frame) between the original SDSS and the follow-up spectroscopy. Using the image magnifications we obtain from the lensing models (Table \ref{tab:models}), and the fact that the source rest frame $i$ band corresponds to the observed $H$ - $Ks$ bands, we estimate the source quasar absolute brightness to be $M_i\sim-22.0$ to $\sim-25.7$, depending on the choice of lensing model. Since the structure function is larger for fainter quasars, we use the former value and obtain an upper estimate of $\Delta m\sim0.23$ (because this is a magnitude difference, it has the same value in the rest frame as in the magnified, observed-frame). This value is too small to explain the spectral changes at short wavelengths. We note however that the structure function might be able to explain the flux changes of $\sim 0.3$ mag ($\Delta m=0.25$) in both quasars in the $z$ band, in $\sim 4$ years, and it may also introduce flux variations of about 0.26 - 0.23 mag at the observed-frame $J$ - $Ks$ bands, during the maximum rest frame time interval of the quasar sample, $\sim 2$ years. We point out however that the structure function is a statistical average over a sample of quasars, therefore brightness changes of larger amplitude are not ruled out in individual objects (magnitude differences larger that 0.75 occur in about 1\% of quasars; \citet{vandenberk04}).

\subsection{Estimates of the intrinsic flux ratio}\label{sec:summaryspec}

\begin{figure}
\epsscale{0.9}
\plotone{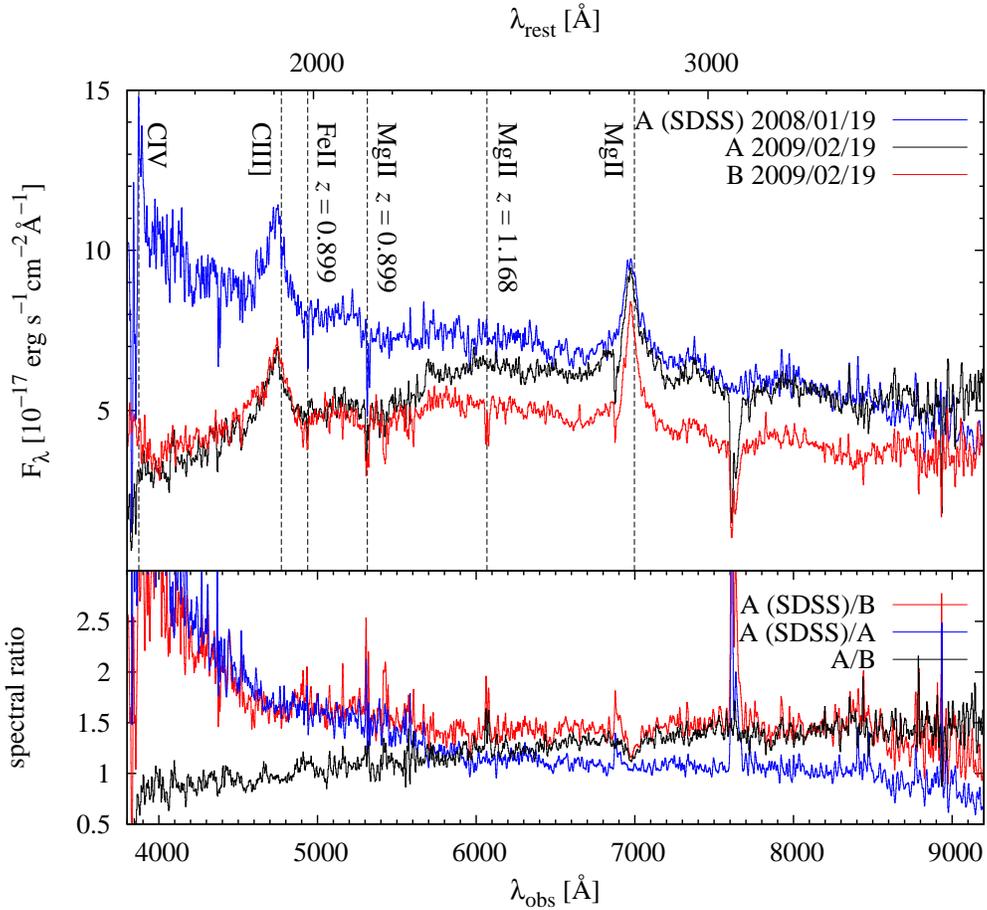}
\caption{\emph{Top:} Spectra of the two stellar components (quasars) in SDSS J1320+1644, A and B. For component A, we plot the original SDSS spectrum (blue), as well as the ARC 3.5 m follow-up spectrum (black). The quasar designation corresponds to Figure \ref{fig:subaru-imag} (object A is the component to the South-West, in Figure \ref{fig:sdss-imag}). The emission and absorption lines we identified are marked. Two absorption systems in the follow-up spectra, at the blue wing of \ion{Mg}{2} and at $\sim7600${\,\AA}, are due to atmospheric absorption (these have been calibrated out of the original SDSS spectrum). \emph{Bottom:} the ratio of all three spectral pairs.
\label{fig:spec}}
\end{figure}

Our analysis shows that the spectral shapes of A \& B (in the continua) are compatible with the gravitational lensing hypothesis, where the differences are caused by differential extinction, microlensing and intrinsic variability. Microlensing and extinction are more important at short wavelengths than intrinsic variability, and slit losses may also contribute to the changes in the spectrum of A. Regarding the intrinsic flux ratio, all three spectral flux ratios of the two quasars, which we plotted in Figure \ref{fig:spec}, are fairly constant in the range 7200 $-$ 8400 \AA\ , with A(SDSS)/A (the original SDSS spectrum of A divided by the follow-up spectrum of A) fairly close to 1, and A/B, A(SDSS)/B $\sim$ 1.4. This latter value is in excellent agreement with the flux ratio we measure in the $J, H$ and $Ks$ bands, where we expect differential extinction and microlensing to be weakest. Together, these results suggest that the intrinsic flux ratio, which would ideally be measured in the far-infrared or radio, is $\sim1.4 \pm 0.2$, and all three flux-altering physical factors considered are weak in the 7200 $-$ 8400 \AA\ spectral range. Here we chose the error bar on the flux ratio somewhat arbitrarily, slightly larger than the scatter in the flux ratio at 7200 $-$ 8400 \AA\, as well as the $J, H$ and $Ks$ bands.

The two quasars were also detected in the mid-infrared by WISE \citep{wright10}, at almost the same epoch as the Subaru MOIRCS $J, H$ and $Ks$ observations. The flux ratio at 3.4, 4.6 and 12 $\mu$m is $\sim1$ (Table \ref{tab:wise}), but we consider this to be broadly consistent with the value above, because the quasars were detected at quite low S/N $\sim10-20$, and the fluxes were measured by deblended profile fits, at very poor angular resolution (6$\farcs$1, 6$\farcs$4, 6$\farcs$5 and 12$\farcs$0 in the four bands, respectively). There is also the possibility of unknown flux contamination from the two bright galaxies. The quasar pair also appears to be radio quiet, based on non-detection in the Faint Images of the Radio Sky at Twenty-cm survey, which is sensitive to $> 1$ mJy/beam. 

In view of the WISE data, we also consider the possibility that the intrinsic flux ratio is $\sim1.0 \pm 0.2$ (here we have broadened the scatter in the values measured at at 3.4, 4.6 and 12 $\mu$m). This requires that the flux ratio of $\sim1.4$ in the continuum, found in the data redward of the \ion{Mg}{2} BEL, is due to microlensing. This assumption is still in agreement with the results in \S\ref{sec:microlensing}, and in particular Equation \ref{eq:tsrc}, which is an order-of-magnitude estimate. In this interpretation, the flux change at long wavelengths occurs on timescales larger than those sampled by our observations, whereas those at shorter wavelengths are noticeable. This assumption is testable via deeper, high resolution mid-infrared observations, as the mid-infrared emitting region is expected to be too large for microlensing to occur. This interpretation would also explain why the flux ratio decreases slightly with increasing wavelength between the $H$ and $Ks$ bands, reaching unity in the WISE bands. 

\section{Environment of the lens}\label{sec:discuss-cluster}

Since our attempts to model the system as a two-image gravitationally lensed quasar require a significant additional dark matter component, we next look at the environment of the lens, in order to search for clues to the existence of a possible galaxy cluster.

We proceed by making a color-color diagram of the galaxies in the MOIRCS field of view ($\sim 1900 \times 1660\ h_{70}^{-1}$ kpc at $z=0.899$, centered on SDSS~J1320+1644), using the $J$, $H$ and $Ks$ bands. We aim to isolate the galaxies that are similar in color to G1 and G2, and study their spatial distribution around the target. We use SExtractor \citep{bertin96} to catalogue and match all the objects in the field of view, and we isolate the galaxies by requiring the SExtractor stellarity parameter to be less than 0.2 in all three bands. We eliminate a few galaxies that are very close together, so that aperture photometry would fail, which leads to a sample of 98 galaxies.

From Table \ref{tab:photometry}, using the aperture photometry, we determine the colors ($J-H$, $H-Ks$) = (0.87 $\pm$ 0.07, 0.88 $\pm$ 0.05) for G1, and (0.87 $\pm$ 0.08, 0.80 $\pm$ 0.05) for G2. In color-color space, G1 and G2 lie close to each other, and close to the concentration at  ($J-H$, $H-Ks$) $\sim$ (0.9, 0.7). Thus there may be an excess of galaxies at the redshift of G1 and G2, indicative of a cluster. 

\begin{figure}
\epsscale{1}
\plotone{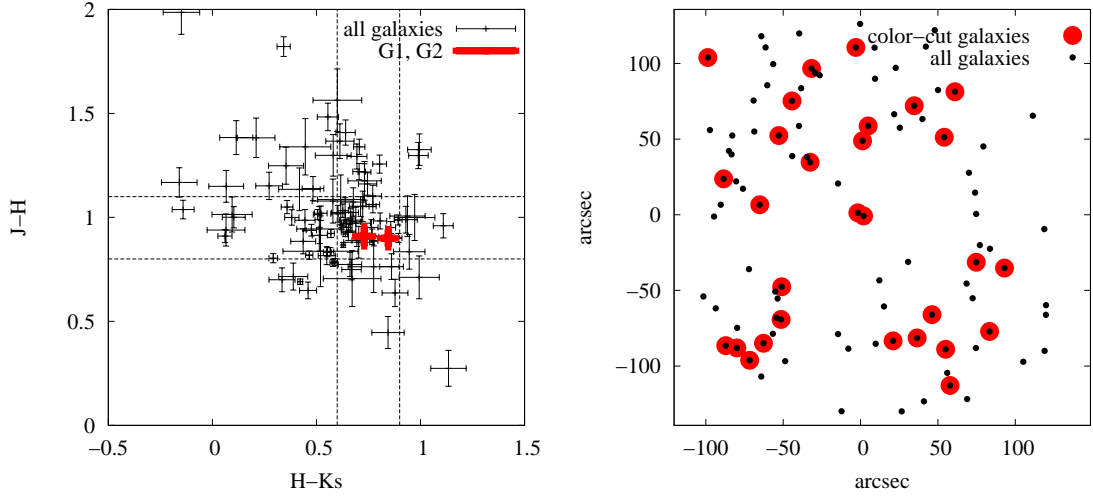}
\caption{Color-color ({\it left}) and spatial distribution diagrams ({\it right}) of the galaxies in the Subaru MOIRCS field of view. The colors and associated error bars are based on aperture photometry (MAG$\_$APERT) determined with SExtractor. The color-cut region used to isolate galaxies possibly associated with the candidate cluster at $z\sim0.9$ is marked with dashed lines. The spatial distribution diagram identifies the galaxies whose colors match the color-cut region.  We use the field of view of only one of the MOIRCS detectors, centered on the target. East is up and North is to the right. 100 arcsec at $z=0.9$ corresponds to a proper scale of 790 $h^{-1}_{70}$ kpc. [\it{See the electronic edition of the Journal for a color version of this figure.}]
\label{fig:colorscluster}}
\end{figure}

We attempt a color cut at 0.8 $< J-H <$ 1.1 and 0.6 $< H-Ks <$ 0.9, which includes the concentration of galaxies as well as G1 and G2. In Figure \ref{fig:colorscluster} (right), we plot the positions of the galaxies selected by this color-cut from the MOIRCS field of view centered on SDSS~J1320+1644; they seem to spatially surround the target. We caution however that the broad color cut may contain galaxies at different redshift. A multidimensional color plot would help break the degeneracies,  but the UH88 bands, even the deep $z$ band, detect only a few of the galaxies visible in the MOIRCS bands and do not allow a systematic treatment.

Our analysis therefore hints that G1 and G2 are embedded at the center of a galaxy cluster, which in turn supports the existence of a heavy dark matter halo associated with the cluster, required by our models in \S\ref{sec:mass}. In \S\ref{sec:discuss-MLratio}, we provide an estimate of the content of dark matter.

\section{Summary and conclusions}\label{sec:summary}

We have reported the discovery as part of the SQLS of SDSS~J1320+1644, a pair of quasars at $z=1.487$ with two galaxies in between, in a cross-like configuration. We have estimated the redshift of the two galaxies as $z=0.899$, based on absorption lines observed in the quasar spectra and consistency with the photometric redshifts of the galaxies.

We used the observational constraints to investigate whether the system is a two-image gravitationally lensed quasar or a binary quasar. We showed that the key to differentiate between the gravitational lens and binary pair hypotheses is whether or not there exists a dark matter halo at the location of the system, capable of boosting the gravitational potential of the galaxies. We estimated the required virial mass or velocity dispersion of this halo, as well as the expected time delays and image magnifications, using a variety of SIS and NFW profiles. We were also able to show that the halo can only be centered on the brighter, elliptical galaxy G1 or somewhere in between the two quasars (but not on the disk-like galaxy G2), as well as to set constraints on the velocity dispersions of these two galaxies, and an additional, fainter third galaxy. The third galaxy may be located at either $z=0.899$ or $z=1.168$. In the first case, it would further support the existence of a halo indicative of a galaxy group or cluster, and in the second it would explain the presence of a strong absorption system in one of the quasars.

We have found a substantial number of galaxies in the field, surrounding the quasar pair, with colors consistent with $z=0.899$ (based however on only three filters). If this is a galaxy cluster, the quasar pair would reside behind the center of the cluster potential, which would provide the required dark matter halo lens. The calculated mass-to-light ratio assuming the lensing hypothesis (\S\ref{sec:discuss-MLratio}) is in agreement with what would be expected from a cluster-scale lens. 

We have performed a comparative analysis of the spectra of the two quasars (as well as inferred the intrinsic flux ratio of the two quasars), and concluded that all observed differences can be attributed to a combination of extinction, microlensing and intrinsic variability, in support of the gravitational lensing hypothesis, with the microlensing being dominant.

The most similar known system to our target is arguably Q0957+561 \citep{walsh79}, which has two quasar images separated by  $6\farcs17$ and one lensing galaxy that is part of a confirmed cluster that boosts the image separation. That system shows, in addition, evidence of strongly lensed arcs, which are the extended images of the lensed quasar host galaxy \citep{bernstein93}.  

We conclude that SDSS~J1320+1644 is a probable gravitationally lensed quasar, although we are unable to prove this beyond doubt, based on our available data. We propose the following observations as a means of establishing the true nature of this object unambiguously:

\begin{enumerate}
  \setlength{\itemsep}{-5pt}
       \item As the most time-effective way of testing the lensing hypothesis, we suggest deep adaptive optics or Hubble Space Telescope observations to study the morphology of the quasar host galaxies, and look for extended arc-like features, characteristic of gravitational lensing. We note that there is  a 16.6 mag (R band) star $\sim58\farcs4$ from the target, which could be used as a tip-tilt star for the Laser Guide Star Adaptive Optics capabilities of Subaru Telescope.
      \item Deep, wide field optical imaging of the region could be used to detect weak lensing associated with the possible lensing cluster. 
      \item Deep X-ray observations would test the presence of the galaxy group/cluster by providing an independent measure of the velocity dispersion of intracluster gas, as well as determining its dynamical center. 
      \item Spectroscopy of the two brightest galaxies would determine if they are indeed at the same redshift, consistent with the spectral absorption lines, as assumed in our galaxy cluster hypothesis. These galaxies are faint in the $V$, $R$, $I$, $z$ and SDSS bands, and therefore the photometric redshifts are fairly uncertain. Multi-object spectroscopy would provide a means to check the redshift of the surrounding galaxies, and therefore either consistently prove or disprove the cluster hypothesis. At the very least, deeper imaging in other bands is required to constrain the photometric redshift of the surrounding galaxies.
  \item Long-time monitoring of the quasar fluxes to look for correlated variations (a time delay) would convincingly reveal the nature of this system, as long as the correlated variations are not suppressed by microlensing. Although we do not possess sufficient constraints to estimate a robust time delay, our best estimates are $\sim 1 - 7$ years, based on our lensing models.
    \end{enumerate} 

Should this system indeed prove to be a gravitational lens, it would be the third-largest separation gravitationally lensed quasar in the SQLS, and the largest separation two-image lensed quasar known. These large separation lenses are important to include in a complete statistical sample such as the one provided by the SQLS \citep{inada12}, in order to constrain the hierarchical structure formation at cluster mass scales. They are also exceedingly rare, because the probability of quasars strongly lensed by clusters is 1-2 orders of magnitudes smaller than that of quasars lensed by galaxies \citep[e.g.,][]{inada12}.

\acknowledgments

This work was supported in part by the FIRST program "Subaru Measurements of Images and Redshifts (SuMIRe)", World Premier International Research Center Initiative (WPI Initiative), MEXT, Japan, and Grant-in-Aid 23740161 and 24740171 for Scientific Research from the Japan Society for the Promotion of Science (JSPS).

The authors would like to thank the anonymous referee for criticism that helped improve the paper. C.E.R. and I.K. acknowledge the support of the JSPS Research Fellowship. D.S. acknowledges supports from Deutsche Forschungsgemeinschaft, reference SL172/1-1. M.A.S acknowledges support from NSF grant AST-0707266. The authors recognize and acknowledge the very significant cultural role and reverence that the summit of Mauna Kea has always had within the indigenous Hawaiian community. We are most fortunate to have the opportunity to conduct observations from this superb mountain. 

This publication makes use of data products from the Wide-field Infrared Survey Explorer, which is a joint project of the University of California, Los Angeles, and the Jet Propulsion Laboratory/California Institute of Technology, funded by the National Aeronautics and Space Administration.

Funding for the SDSS and SDSS-II has been provided by the Alfred
P. Sloan Foundation, the Participating Institutions, the National
Science Foundation, the U.S. Department of Energy, the National
Aeronautics and Space Administration, the Japanese Monbukagakusho, the
Max Planck Society, and the Higher Education  Funding Council for
England. The SDSS Web Site is http://www.sdss.org/. 

The SDSS is managed by the Astrophysical Research Consortium for the
Participating Institutions. The Participating Institutions are the
American Museum of Natural History, Astrophysical Institute Potsdam,
University of Basel, Cambridge University, Case Western Reserve
University, University of Chicago, Drexel University, Fermilab, the
Institute for Advanced Study, the Japan Participation Group, Johns
Hopkins University, the Joint Institute for Nuclear Astrophysics, the
Kavli Institute for Particle Astrophysics and Cosmology, the Korean
Scientist Group, the Chinese Academy of Sciences (LAMOST), Los Alamos
National Laboratory, the Max-Planck-Institute for Astronomy (MPIA),
the Max-Planck-Institute for Astrophysics (MPA), New Mexico State
University, Ohio State University, University of Pittsburgh,
University of Portsmouth, Princeton University, the United States
Naval Observatory, and the University of Washington.

\clearpage

\begin{deluxetable}{cccccc}
\tablewidth{0pt}
\tabletypesize{\footnotesize}
\tablecaption{Original SDSS photometry}
\tablewidth{0pt}
\tablehead{\colhead{Object \& designation} & 
\colhead{$u$} &
\colhead{$g$} &
\colhead{$r$} &
\colhead{$i$} &
\colhead{$z$} }
\startdata
SDSS~J132059.17+164402.59 (A) & $20.16\pm$0.04 & $19.58\pm$0.02 & $19.16\pm$0.02 & $18.88\pm$0.02 & $18.86\pm$0.04 \\
& $20.10\pm$0.06 & $19.54\pm$0.02 & $19.17\pm$0.02 & $18.84\pm$0.02 & $18.84\pm$0.06 \\
    \hline
SDSS~J132059.73+164405.6 (B) &  $18.93\pm$0.02  & $18.84\pm$0.02 & $18.74\pm$0.02 & $18.62\pm$0.02 & $18.67\pm$0.03 \\
&  $18.91\pm$0.02  & $18.78\pm$0.01 & $18.74\pm$0.02 & $18.57\pm$0.02 & $18.69\pm$0.06 \\
    \hline
SDSS~J132059.37+164406.66 (G2) &  $25.05\pm$0.79  & $24.13\pm$0.36 & $23.01\pm$0.22 & $22.10\pm$0.17 & $21.77\pm$0.47 \\
&  $24.53\pm$2.84  & $23.99\pm$0.58 & $23.08\pm$0.41 & $21.96\pm$0.27 & $21.60\pm$0.80
\enddata
\tablecomments{The AB magnitudes are corrected for Galactic extinction following \citet[][]{schlegel98}. The object designations are those from \S\ref{sec:image}. The first line for each object gives the PSF magnitudes (A, B), or best model magnitudes (G2), and the second line gives aperture photometry magnitudes. Objects G3 and G5 (see Figure \ref{fig:subaru-imag}) were also identified as photometric targets by the SDSS pipeline, but their magnitudes are not reproduced, as they are considered unreliable.}
\label{tab:sdss-phot}
\end{deluxetable}


\begin{deluxetable}{lcccccc}
\tablewidth{0pt}
\tabletypesize{\footnotesize}
\tablecaption{Summary of follow-up imaging observations}
\tablewidth{0pt}
\tablehead{\colhead{Filter} &
\colhead{Instrument} & 
\colhead{Exposure} &
\colhead{Airmass} &
\colhead{Seeing} &
\colhead{Zero-point uncertainty} &
\colhead{Observation date}}
\startdata
$V$ & UH88/Tek2k & $3\times100$ s  & 1.20 & $1.0 - 1.2$ & 0.10 & 2009 April 15 \\
$R$ & UH88/Tek2k & $4\times100$ s  & 1.40 & $1.0 - 1.2$ & 0.08 & 2009 April 15 \\
$I$ & UH88/Tek2k & $7\times100$ s  & 1.20 & $1.0 - 1.2$ & 0.07  & 2009 April 15 \\
$z$ & UH88/Tek2k & $8\times100$ s  & 1.20 & $1.0 - 1.2$ & 0.07 & 2009 April 15 \\
$z$ & UH88/WFGS2 & $24\times180$ s  & $1.14 - 1.53$ & $1.0 - 1.2$ & 0.04 & 2009 April 17 \\
$J$ & Subaru/MOIRCS & $4\times150$ s  & 1.03 & 0.4 & 0.01 & 2010 April 2 \\
$H$ & Subaru/MOIRCS & $4\times120$ s  & 1.05 & 0.4 & 0.01 & 2010 April 2 \\
$Ks$ & Subaru/MOIRCS & $4\times150$ s  & 1.09 & 0.4 & 0.01 & 2010 April 2
\enddata
\tablecomments{Zero-point uncertainties in the $V$, $R$, $I$ and $z$ filters include the scatter between the values obtained from the 10 stars used in the estimation, as well as the Lupton (2005) transformations. In the $J$, $H$ and $Ks$ filters, the uncertainties include those quoted in the catalogue and the aperture photometry errors. There the actual error may be $\sim0.05$ mag, if the differences between catalogues and the values obtained from different frames of the standard star are included. Among the two $z$ band observations, only the deeper one is considered throughout this article.}   
\label{tab:followup-data}
\end{deluxetable}


\begin{deluxetable}{lcc}
\tablewidth{0pt}
\tabletypesize{\footnotesize}
\tablecaption{Astrometry results for SDSS~J1320+1644}
\tablewidth{0pt}
\tablehead{\colhead{Object} & 
\colhead{$\Delta$x [arcsec]} &
\colhead{$\Delta$y [arcsec]}}
\startdata
  A & 0.000 $\pm$ 0.002 & 0.000 $\pm$ 0.002 \\
  B & $-8.066$ $\pm$ 0.002 & 2.939 $\pm$ 0.002 \\
  G1 & $-4.991$ $\pm$ 0.006  & 0.117 $\pm$ 0.006 \\
  G2 & $-2.960$ $\pm$ 0.006 & 3.843 $\pm$ 0.007 \\
  G4 & $-9.169$ $\pm$ 0.019 & 5.173 $\pm$ 0.024 \\
  G3 & $-1.682$ $\pm$ 0.018 & $-5.166$ $\pm$ 0.022 \\
  G5 & $-12.177$ $\pm$ 0.021 & 7.176 $\pm$ 0.020
\enddata
\tablecomments{Astrometry of the SDSS J1320+1644 system, determined in the MOIRCS Ks band (the S\'{e}rsic index of the galaxies has been left unconstrained in the GALFIT modeling). The positive directions of x and y are defined towards West and North, respectively. The quoted errors are the GALFIT statistical errors (A, B, G1, G2), and those determined with the IRAF PHOT task (G3, G4, G5).}   
\label{tab:astrometry}
\end{deluxetable}


\begin{deluxetable}{lccccccc}
\tablewidth{0pt}
\tabletypesize{\footnotesize}
\tablecaption{Follow-up photometry of the SDSS J1320+1644 system}
\tablewidth{0pt}
\tablehead{\colhead{Object} & 
\colhead{$V$} &
\colhead{$R$} &
\colhead{$I$} &
\colhead{$z$} &
\colhead{$J$} &
\colhead{$H$} &
\colhead{$Ks$} }
\startdata
  \multirow{2}{*}{A} & 19.23 $\pm$ 0.01 & 18.68 $\pm$ 0.01 & 18.33 $\pm$ 0.01 & 18.62 $\pm$ 0.01 & 17.58 $\pm$ 0.01 & 16.79 $\pm$ 0.01 & 16.62 $\pm$ 0.01 \\
  & 19.20 $\pm$ 0.01 & 18.71 $\pm$ 0.01 & 18.31 $\pm$ 0.01 & 18.71 $\pm$ 0.01 & 17.62 $\pm$ 0.02 & 16.86 $\pm$ 0.01 & 16.67 $\pm$ 0.01 \\
    \hline
   \multirow{2}{*}{B} & 19.30 $\pm$ 0.01 & 18.92 $\pm$ 0.01 & 18.67 $\pm$ 0.01 & 19.00 $\pm$ 0.01 & 18.05 $\pm$ 0.01 & 17.29 $\pm$ 0.01 & 17.03 $\pm$ 0.01 \\
  & 19.28 $\pm$ 0.01 & 18.94 $\pm$ 0.01 & 18.66 $\pm$ 0.01 & 19.10 $\pm$ 0.02 & 18.04 $\pm$ 0.02 & 17.29 $\pm$ 0.02 & 16.98 $\pm$ 0.02 \\
    \hline
   \multirow{2}{*}{G1} & 23.63 $\pm$ 0.37 & 23.01 $\pm$ 0.22 & 21.66 $\pm$ 0.11 & 21.62 $\pm$ 0.10 & 19.96 $\pm$ 0.02 & 18.83 $\pm$ 0.11 & 18.13 $\pm$ 0.04 \\
   & 23.02 $\pm$ 0.23 & 22.85 $\pm$ 0.16 & 21.47 $\pm$ 0.09 & 21.83  $\pm$ 0.12 & 20.00 $\pm$ 0.06 & 19.13 $\pm$ 0.04 & 18.25 $\pm$ 0.03 \\
      \hline
    \multirow{2}{*}{G2} & 23.16 $\pm$ 0.24 & 23.04 $\pm$ 0.22 & 22.12 $\pm$ 0.16 & 21.51 $\pm$ 0.08 & 20.14 $\pm$ 0.02 & 19.24 $\pm$ 0.02 & 18.50 $\pm$ 0.01 \\
   & 23.25 $\pm$ 0.28 & 23.52 $\pm$ 0.29 & 22.52 $\pm$ 0.23 & 21.65 $\pm$ 0.10 & 20.20  $\pm$ 0.06 & 19.33 $\pm$ 0.05 & 18.54 $\pm$ 0.04 \\
   \hline
   G3 & & & & & 21.42 $\pm$ 0.12 & 20.24 $\pm$ 0.10 & 20.07 $\pm$ 0.10 \\
   \hline
    G4 & & & & & 21.86 $\pm$ 0.14 & 21.05 $\pm$ 0.14 & 20.35 $\pm$ 0.10 \\
    \hline
   G5 & & & & & 22.23 $\pm$ 0.17 & 21.08 $\pm$ 0.14 & 20.80 $\pm$ 0.13
\enddata
\tablecomments{The values are in magnitudes (all in the Vega system, except for the $z$ band, in the AB system). For A, B, G1 and G2, the first line indicates model magnitudes provided by GALFIT (morphological fit), and the second line aperture photometry obtained with the IRAF PHOT task. For G3, G4 and G5, only aperture photometry results are given. Galactic extinction \citep{schlegel98} and atmospheric extinction have been corrected. Quoted errors are statistical errors only. Errors do not include uncertainties in the photometric zero-points (Table \ref{tab:followup-data}), in the PSF or due to the galaxy modeling parameters. Aperture photometry has been measured in all bands with a fixed  $\sim3\farcs1$ diameter aperture ($\sim1\farcs9$ for G4 and G5), selected to maximize the amount of light from the target, as well as minimize contamination from the other objects. In the $V$, $R$, $I$ and $z$ bands, contamination may be higher due to the larger seeing. In order for the models to converge, the morphological parameters of the S\'{e}rsic profiles fitted to these four bands were constrained to the values determined in the $Ks$ band (i.e., S\'{e}rsic index fixed to the closest canonical value of 1 or 4, effective radius).}   
\label{tab:photometry}
\end{deluxetable}


\begin{deluxetable}{lccccc}
\tablewidth{0pt}
\tabletypesize{\footnotesize}
\tablecaption{Morphological parameters of the two bright galaxies}
\tablewidth{0pt}
\tablehead{\colhead{Object} & 
\colhead{Effective radius [arcsec]} &
\colhead{S\'{e}rsic index} &
\colhead{Axis ratio (b/a)} &
\colhead{Position angle [deg]}}
\startdata
G1 & 0.24$\pm$0.01 & 4.72$\pm$0.76 & 0.60$\pm$0.04 & $-$3.6$\pm$3.5 \\
G2 & 0.37$\pm$0.01 & 0.57$\pm$0.07 & 0.66$\pm$0.02 & 2.6$\pm$2.7
\enddata
\tablecomments{Morphological parameters are fitted by GALFIT in the $Ks$ band, with attached statistical errors. The position angle is measured from North towards East.}   
\label{tab:morphology}
\end{deluxetable}


\begin{deluxetable}{lcccccc}
\tablewidth{0pt}
\tabletypesize{\footnotesize}
\tablecaption{Best-fit photometric redshifts}
\tablewidth{0pt}
\tablehead{\colhead{Object \& template} & 
\colhead{Used filters $z$} &
\colhead{Best-fit $z$} &
\colhead{$1\sigma$ limits} &
\colhead{$2\sigma$ limits} &
\colhead{$\chi^2/_{\mbox{d.o.f.}}$ (d.o.f.)} }
\startdata
G1, E (EAzY) & VRI$z$JHKs & $0.81$  & $0.67 - 0.96$ & $0.53 - 1.10$ & 0.34 (6) \\
G1, E (HyperZ) & VRI$z$JHKs  & 0.87 & $0.77 - 0.94$ & $0.71 - 0.98$ & 0.65 (6) \\
G1, E (EAzY) & JHKs & $0.83$  & $0.65 - 1.02$ & $0.38 - 1.20$ & 0.17 (2) \\
G1, E (HyperZ) & JHKs & $0.93$  & $0.79 - 1.04$ & $0.71 - 1.19$ & 0.03 (2) \\
G2, Sbc (EAzY) & $ugri$VRI$z$JHKs & $0.80$  & $0.71 - 0.89$ & $0.64 - 0.97$ & 0.46 (10)\\
G2, Sbc (HyperZ) & $ugri$VRI$z$JHKs & $0.86$  & $0.80 - 0.91$ & $0.67 - 0.95$ & 1.97 (10)\\
G2, Sbc (EAzY) & JHKs & $0.79$  & $0.57 - 0.97$ & $0.38 - 1.32$ & 0.29 (2) \\
G4, all (EAzY) & JHKs & $1.11$  & $0.56 - 1.58$ & $0.29-1.98$ & 0.05 (2) \\
G3, all (EAzY) & JHKs & $0.55$  & $0.47 - 1.10$ & $0.27 - 1.55$ & 2.07 (2) \\
G5, all (EAzY) & JHKs & $0.60$  & $0.52 - 1.45$ & $0.30 - 1.94$ & 0.84 (2)
\enddata
\label{tab:photoz}
\end{deluxetable}


\begin{deluxetable}{lccccc}
\tablewidth{0pt}
\tabletypesize{\footnotesize}
\tablecaption{Parameters of the best-fit lensing models}
\tablewidth{0pt}
\tablehead{\colhead{model} & 
\colhead{$\sigma$ [km s$^{-1}$] or M [$h_{70}^{-1} M_\odot$]} &
\colhead{$e$ or $\gamma$} &
\colhead{$\theta_e$ or $\theta_\gamma$ [deg]} &
\colhead{$\mu_{\rm tot}$} &
\colhead{$\Delta t$ [days]}}
\startdata
SIS$+\gamma$ at G1 & $710 \pm 8$ km s$^{-1}$ & $0.026 \pm 0.011$ & $20 \pm 20$ & $18 \pm 2$ & $911 \pm 21$ \\
  \hline
\multirow{2}{*} {SIS free} & \multirow{2}{*} {$645 \pm 25$ km s$^{-1}$} & \multirow{2}{*}  -  & \multirow{2}{*}  -  & \multirow{2}{*} {$37 \pm 29$} & $-860 \pm 460$ \ /\  \\ 
 & & & & & $1630 \pm 940$\\ 
 \hline
\multirow{3}{*}{elliptical NFW at G1} & $(5.5 \pm 0.5) \times10^{13}\ h_{70}^{-1} M_\odot$ ($c_{\rm vir}=30)$ & $0.03 \pm 0.01$ & $20 \pm 13$ & $13.5 \pm 0.5$ & $1095 \pm 15$ \\
& $(4.2 \pm 0.7)\times10^{14}\ h_{70}^{-1} M_\odot$ ($c_{\rm vir}=6$) & $0.018 \pm 0.012$ & $\sim 0 - 180$ & $60 \pm 12$ & $510 \pm 40$ \\ 
& $(3.15 \pm 0.35)\times10^{15}\ h_{70}^{-1} M_\odot$ ($c_{\rm vir}=2$) & $0.01 \pm 0.01$ & $\sim 0 - 180$ & $147 \pm 43$ & $320 \pm 41$\\ 
  \hline
\multirow{3}{*}{NFW free} & $(4.9 \pm 0.4)\times10^{13}\ h_{70}^{-1} M_\odot$ ($c_{\rm vir}=30$) & - & - & $41 \pm 18$ & $-638 \pm 262$ \\
& $(4.0 \pm 0.4)\times10^{14}\ h_{70}^{-1} M_\odot$ ($c_{\rm vir}=6$) & - & - & $99 \pm 46$ & $-547 \pm 223$ \\ 
& $(3.2 \pm 0.1)\times10^{15}\ h_{70}^{-1} M_\odot$ ($c_{\rm vir}=2$) & - & - & $175 \pm 75$ & $-380 \pm 80$
\enddata
\tablecomments{The range on each parameter is obtained by changing the velocity dispersions of G1, G2 and G4 over the allowed $1\sigma$ range (see \S\ref{sec:mass}). $e$ and $\gamma$ are the ellipticity and shear, respectively, and $\theta_e$, $\theta_\gamma$ are the respective position angles (measured East of North). $\mu_{\rm tot}$ is the total magnification of images A and B, and the predicted time delay $\Delta t$ is positive if image A leads.}
\label{tab:models}
\end{deluxetable}

\begin{deluxetable}{ccccc}
\tablewidth{0pt}
\tabletypesize{\footnotesize}
\tablecaption{WISE mid-infrared photometry of the SDSS~J1320+1644 quasar pair}
\tablewidth{0pt}
\tablehead{\colhead{Object \& flux ratio} & 
\colhead{W1 (3.4 $\mu$m)} &
\colhead{W2 (4.6 $\mu$m)} &
\colhead{W3 (12 $\mu$m)} &
\colhead{W4 (22 $\mu$m)}}
\startdata
A & $15.32\pm0.05$ & $13.99\pm0.05$ & $10.92\pm0.10$ & $9.06\pm0.53$ \\
B & $15.32\pm0.05$ & $14.04\pm0.05$ & $10.85\pm0.09$ & $7.97\pm0.20$ \\
A/B & $1.00\pm0.07$ & $1.05\pm0.07$ & $0.94\pm0.12$ & $0.53\pm0.19$
\enddata   
\tablecomments{The deblended profile-fit magnitudes are in the Vega system. The observations were performed between 2010, January 7 and 2010, June 30. The values in W4 are particularly unreliable, as the quasars are detected at S/N $\sim2-5$, with an angular resolution of 12$\farcs$0, larger than the quasar separation.} 
\label{tab:wise}
\end{deluxetable}

\clearpage

\appendix

\section{Additional considerations on the lens modeling}\label{sec:appen-lens}

\subsection{A lens model for G4}\label{sec:appen-G4}

By considering that G4 is located at $z=1.168$ and applying the F-J and T-F laws, the velocity dispersion of this galaxy is $\sigma=175\pm40$ km s$^{-1}$ or $125\pm12$ km s$^{-1}$, respectively. Assuming that the effect of G4 on the lensing configuration is small, we can treat G4 as an SIS lens situated at $z=0.899$, but with a smaller effective velocity dispersion. This eliminates the need to introduce multiple lens planes and is justified by the fact that the convergence from G4 at B (arbitrarily taken as the object closest to G4, but projected to $z=1.168$) is as small as $\kappa\sim \theta_{\rm Ein}/2d_{B-G4}\sim 0.028 \pm 0.012$ (F-J) or $\sim 0.014 \pm 0.002$ (T-F), where $\theta_{\rm Ein}$ is the Einstein radius of the lens G4 estimated from the velocity dispersion (assuming an SIS profile) and $d_{B-G4}=2\farcs49$ is the projected angular distance between B and G4. From \citet[][]{keeton03}, using $\kappa$ (convergence) = $\gamma$ (shear) for the SIS profile, the effective convergence of G4 at $z=0.899$ is $\kappa_{\rm eff}=(1-\beta)\kappa/(1-2\beta\kappa)\sim 0.012 \pm 0.005$ (F-J), or $\sim 0.006 \pm 0.002$ (T-F), where $\beta=D_{12}D_{\rm OS}/(D_{\rm O2}D_{\rm 1S})$, $D_{ij}=D(z_i, z_j)$ are angular diameter distances, $z_1=0.899, z_2=1.168$, and O, S refer to the observer and source, respectively. The effective convergence corresponds to an SIS profile at G4 with effective velocity dispersion $\sim 81 \pm 19$ km s$^{-1}$ (F-J) or $\sim 57 \pm 5$ km s$^{-1}$ (T-F). We conclude that, considering all possibilities, an SIS model at $z=0.899$ for G4 would have a velocity dispersion in the $1\sigma$ range of $\sim 118 \pm 66$ km s$^{-1}$.

\subsection{Dependence on $c_{vir}$}\label{sec:appen-cvir}

\begin{figure}
\epsscale{0.7}
\plotone{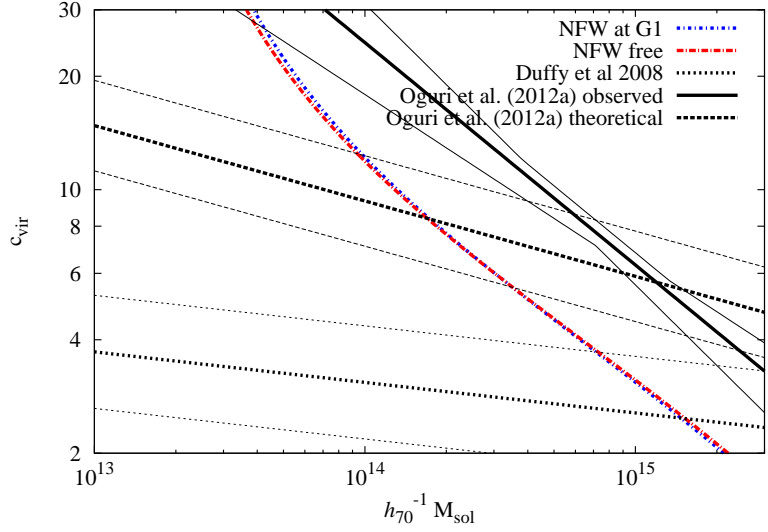}
\caption{The $c_{vir}$ - virial mass degeneracy of the NFW profiles. The thick black lines show published relations from both simulations and observations, while the thin lines show the corresponding 1$\sigma$ ranges. A caveat is that the \citet{oguri12-1} theoretical curve has been derived for a cluster redshift $z=0.45$, which corresponds to the mean redshift of their observed cluster sample. The blue and red curves show the results of these work, for the NFW profile fixed at G1, as well as with unconstrained (free) location, respectively. [\it{See the electronic edition of the Journal for a color version of this figure.}]
\label{fig:cM}}
\end{figure}

Here we investigate how the choice of the concentration parameter $c_{vir}$ of the NFW profiles affects our results in \S\ref{sec:mass}. We explore $c_{vir}$ in the range from 2 to 30, and we optimize all other model parameters. Increasing $c_{vir}$ has the effect of decreasing the total magnification of the images, and increasing the time delay (Table \ref{tab:models}). We plot the resulting $c_{vir}$ - virial mass relation in Figure \ref{fig:cM}. We obtain virtually the same degeneracy for the free NFW profile as for the one fixed at G2. For comparison, we also plot some recently published results on the $c_{vir}$ - virial mass relation:  the results of $N-$body simulations by \citet{duffy08}, as well as the theoretical and observational results inferred by \citet{oguri12-1} from fitting lensing clusters at $z \sim 0.45$. The strong lens selected sample behind the \citet{oguri12-1} relations is known to be biased towards overestimating the concentrations due to halo triaxiality. If SDSS~J1320+1644 is indeed a lens system, it should include the same bias, as it is selected as a strong lens. All these relations are observed to intersect the SDSS~J1320+1644 curves. 

\subsection{Caveats on the velocity dispersions of the galaxies}\label{sec:appen-caveat}

We discuss two caveats. A first caveat is that, when estimating the range of velocity dispersion for each galaxy in \S\ref{sec:mass}, we have not taken into account the possible evolution of the T-F and F-J with redshift. Although there is no consensus in the literature in this regard (e.g., \citet{bohm04} \& \citet{fernandez10} for T-F, \citet{rusin03} \& \citet{treu06} for F-J), the tendency is that galaxies had smaller rotational velocities and velocity dispersions in the past for the same luminosity. This would reduce even more the contribution of the galaxies to the mass models dominated by the dark matter halo. 

A second caveat applies to G4, in the case that this galaxy is located at $z=1.168$. In this case, G4 would also be lensed by the dark matter halo at $z=0.899$, and therefore magnified by a factor $\sim 1.5-3$, depending on the assumed lensing model. This would mean that the galaxy is intrinsically fainter, and its contribution to the mass models would be further reduced. On the other hand, the undeflected position of G4 would be closer to the main deflector, increasing its influence. We have checked that the best-fit models considered in this section (from which we exclude the contribution of G4) would not produce multiple images of this galaxy ($z=1.168$ being closer to the redshift of the main lens than $z=1.487$, the critical surface density for multiple images is increased).  

\section{Mass-to-light ratios}\label{sec:discuss-MLratio}

Here we compute the mass-to-light (M/L) ratio by assuming that SDSS~J1320+1644 is a lensed system, and compare it with the M/L ratio of the other lenses discovered by the SQLS\footnote{\url{http://www-utap.phys.s.u-tokyo.ac.jp/~sdss/sqls/lens.html}} \citep[][]{inada12}. We fit each lens with an SIS profile, in which case the Einstein radius $\theta_E$ is simply half the image separation. For the quasars with more than two images, we consider $\theta_E$ to be half the largest separation between the images. The mass inside the Einstein radius is simply determined as

\begin{equation}
M(\leqslant\theta_E)=\frac{D_\mathrm{OL}D_\mathrm{OS}}{D_\mathrm{LS}}\frac{c^2\theta_E^2}{4G}\simeq 1.23 \times 10^8\  \ h_{70}^{-1}\  \mbox{M}_\odot \left(\frac{D_\mathrm{OL}D_\mathrm{OS}/D_\mathrm{LS}}{1\ h_{70}^{-1}\ \mbox{Mpc}}\right)\ \left(\frac{\theta_\mathrm{E}}{1\ \mbox{arcsec}}\right)^2.
\label{eq:lensmass}
\end{equation}

To determine the luminosity of the SQLS lenses, we referred to the discovery papers \citep[][and references therein]{inada12} and estimated the rest frame R band magnitudes of all the galaxies with available photometry, located between the images, using the HyperZ algorithm.  To convert the rest frame magnitudes into luminosities we used $M_R = 4.48 - 2.5 \log L/L_{\odot,R}$, where $M_{\odot,R} = 4.48$ is the R band magnitude of the Sun. The M/L ratios were computed in terms of the solar M/L ratio, $\Upsilon_\odot$ = 1 M$_\odot$/L$_\odot$.

We attached no error bars to the results of these rough calculations. Not accounting for the effects of lens ellipticity and external shear for each (mostly double-image) system should introduce an error of $\sim10\%$ on the masses \citep[][]{rusin05}. Also, the redshifts of several lenses are not accurately known, which influences both the masses and the inferred luminosities. Photometry for most lenses is only available in three bands (we eliminate 10 objects for which the lens is detected in only one band), and their morphology is typically unknown. It is therefore difficult to choose the appropriate CWW spectral template for calculating the rest frame $R$ band luminosity. However we know that the majority of lens galaxies are ellipticals, as these have larger velocity dispersions \citep[][]{turner84}, and so we generally assume the elliptical template, unless the lens is known to be a spiral, or the fit of another template is much better. Also, luminosities should be more accurate for lenses at $z \sim$ 0.2 - 0.3, because at this redshift the rest frame R band is redshifted into the observed-frame $I$ band, where observations are usually available. For SDSS J1320+1644, the lens redshift estimate means that the rest frame $R$ band is redshifted into the J band, where photometry is available.

\begin{figure}
\epsscale{0.7}
\plotone{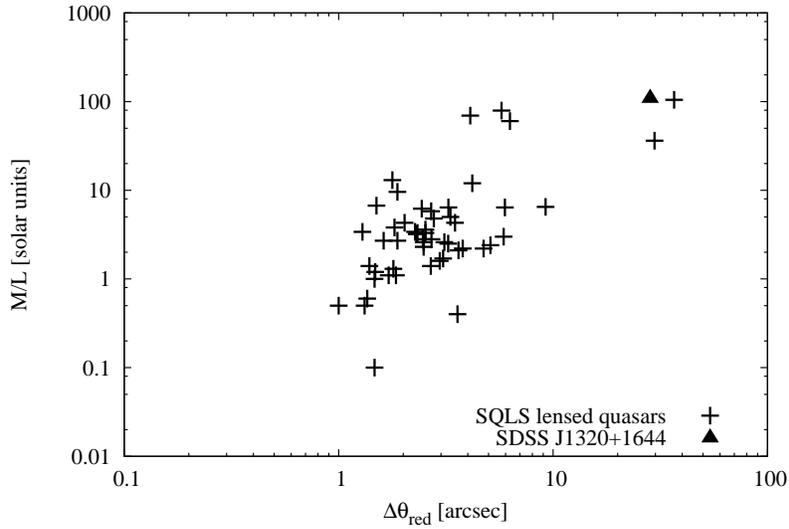}
\caption{Mass-to-light ratios in the rest frame R band for 52 SQLS gravitationally lensed quasars, as well as for SDSS~J1320+1644 (assuming it is also a lensed quasar). Here $\Delta\theta_{red}$ is the reduced image separation (see \S\ref{sec:discuss-MLratio}).
\label{fig:M/L}}
\end{figure}

The results of our M/L estimates are shown in Figure \ref{fig:M/L}, in which we plot the mass-to-light ratios against the reduced image separations. The reduced image separation $\Delta\theta_\mathrm{red}\equiv\Delta\theta\ 
\cdot 
D_\mathrm{OS}/D_\mathrm{LS}$ scales out the dependence on the source and lens distance, thus representing a physical property of the lensing object.

The image separation is the most important observable, because it reflects the depth of the gravitational potential of the lensing object and therefore the structure responsible for the lensing phenomenon. Since clusters are dominated by dark matter much more than normal galaxies, we expect that, as the image separation increases from the galaxy lens to the cluster lens range, so does the M/L ratio. Accordingly, the result we obtain in Figure \ref{fig:M/L} shows a positive correlation between the image separation and M/L. Most small-separation lenses at $\Delta\theta\sim1"$ are clustered at M/L $\sim1-10 \ \Upsilon_\odot$, consistent with the estimates from the dynamics of elliptical galaxies and their stellar populations \citep[][]{padmanabhan04}, whereas the two known cluster-scale lenses SDSS~J1004+4112 and SDSS~J1029+2623 both have a larger M/L. We obtain a similarly large M/L for SDSS J1320+1644.

\clearpage

\clearpage


\begin{thebibliography}{}

\bibitem[Abazajian et al.(2003)]{abazajian03}
Abazajian, K., et al.\ 2003, \aj, 126, 2081

\bibitem[Abazajian et al.(2004)]{abazajian04}
Abazajian, K., et al.\ 2004, \aj, 128, 502

\bibitem[Abazajian et al.(2005)]{abazajian05}
Abazajian, K., et al.\ 2005, \aj, 129, 1755

\bibitem[Abazajian et al.(2009)]{abazajian09}
Abazajian, K.~N., et al.\ 2009, \apjs, 182, 543 

\bibitem[Adelman-McCarthy et al.(2006)]{adelman06}
Adelman-McCarthy, J.~K., et al.\ 2006, \apjs, 162, 38

\bibitem[Adelman-McCarthy et al.(2007)]{adelman07}
Adelman-McCarthy, J.~K., et al.\ 2007, \apjs, 172, 634

\bibitem[Adelman-McCarthy et al.(2008)]{adelman08} 
Adelman-McCarthy, J.~K., et al.\ 2008, \apjs, 175, 297 

\bibitem[Auger et al.(2009)]{auger09}
Auger, M.~W., Treu, T., Bolton, A.~S., Gavazzi, R., Koopmans, L.,~V,~E., Marshall, P.~J., Bundy, K., Moustakas, L.~A.\ 2009, \apj, 705, 1099

\bibitem[Benitez(2000)]{benitez00}
Benitez, N.\ 2000, \apj, 536, 571

\bibitem[Bernstein et al.(1993)]{bernstein93}
Bernstein, G.~M., Tyson, J.~A. \& Kochanek, C.~S.\ 1993, \apj, 105, 816

\bibitem[Bertin \& Arnouts(1996)]{bertin96}
Bertin, E., \& Arnouts, S.\ 1996, AAPS, 117, 393

\bibitem[Binney \& Tremaine(2008)]{binney08}
Binney, J., \& Tremaine, S. 2008, Galactic Dynamics: Second Edition, ed. Binney, J. \& Tremaine, S. (Princeton University Press)

\bibitem[B\"{o}hm et al.(2004)]{bohm04}
B\"{o}hm, A., et al.\ 2004 \aap, 420, 97 

\bibitem[Bolzonella et al.(2000)]{bolzonella03}
Bolzonella, M., Miralles, J.-M., \& Pell{\'o}, R.\ 2000, \aap, 363, 476 

\bibitem[Blanton et al.(2003)]{blanton03}
Blanton, M.~R., Lin, H., Lupton, R.~H., Maley, F.~M., 
Young, N., Zehavi, I., \& Loveday, J.\ 2003, \aj, 125, 2276 

\bibitem[Brammer et al.(2008)]{brammer08}
Brammer, G.~B., van Dokkum, P.~G., \& Coppi, P.\ 2008, \apj, 686, 1503 

\bibitem[Bruzual \& Charlot(1993)]{bruzual93}
Bruzual A., G., \& Charlot, S.\ 1993, \apj, 405, 538 

\bibitem[Chang \& Refsdal(1979)]{chang79}
Chang, K., \& Refsdal, S.\ 1979, \nat, 282, 561


\bibitem[Churchill et al.(2005)]{churchill05}
Churchill, C.~W., Kacprzak, G.~G., Steidel,C.~C.\ 2005, in Proceedings IAU Colloquium No. 199, \ion{Mg}{2} absorption through intermediate redshift galaxies

\bibitem[Claeskens \& Surdej(2002)]{claeskens02}
Claeskens, J.-F., \& Surdej, J.\ 2002, A\&AR, 10, 263 

\bibitem[Clavel et al.(1991)]{clavel91}
Clavel, J. et al.\ 1991, \apj, 366, 64

\bibitem[Coleman et al.(1980)]{coleman80}
Coleman, G.~D., Wu, C.-C., \& Weedman, D.~W.\ 1980, \apjs, 43, 393 

\bibitem[Cutri et al.(2012)]{cutri12}
Cutri, R.~M., et al.\ 2012, Explanatory Supplement to the WISE All-Sky Data Release Products, \url{http://www.apo.nmsu.edu/\\ 35m\_operations/35m\_manual/Instruments/DIS/DIS.html} 

\bibitem[Decarli et al.(2009)]{decarli09}
Decarli, R., Treves, A., \& Falomo, R.\ 2009, \mnras, 396, L31 

\bibitem[Djorgovski(1991)]{djorgovski91}
Djorgovski, S.\ 1991, in Crampton D., ed., ASP Conf. Ser. Vol. 21, The Space Distribution of Quasars. Astron. Soc. Pac., San Francisco, p. 349

\bibitem[Doi et al.(2010)]{doi10}
Doi, M., et al.\ 2010, \aj, 139, 1628

\bibitem[Duffy et al.(2008)]{duffy08}
Duffy, A.~R., Schaye, J., Kay, T.~J., Dalla Vecchia, C. \ 2008, \mnras, 390, L64

\bibitem[Faber \& Jackson(1976)]{faber76}
Faber, S.~M., \& Jackson, R.~E.\ 1976, \apj, 204, 668

\bibitem[Falco et al.(1999)]{falco99}
Falco, E.~E., et al.\ 1999, \apj, 523, 617

\bibitem[Faure et al.(2008)]{faure08}
Faure, C., et al.\ 2008, \apjs, 176, 19


\bibitem[F\'{e}rnandez Lorenzo(2010)]{fernandez10}
F\'{e}rnandez Lorenzo, M., Cepa, J., Bongiovanni, A., P\'{e}rez Garc\'{i}a, A.~M., Lara-L\'{o}pez, M.~A., Popov\'{i}c, M., \& S\'{a}nchez-Portal, M.\ 2010, \aap, 521, 27

\bibitem[Fukugita et al.(1996)]{fukugita96}
Fukugita, M., Ichikawa, T., Gunn, J.~E., Doi, M., Shimasaku, K., 
\& Schneider, D.~P.\ 1996, \aj, 111, 1748


\bibitem[Green et al.(2010)]{green10}
Green, P.~J., Myers, A.~D., Barkhouse, W., A., Mulchaey, J.~S., Bennert, V.~N., Cox, T., J., Aldcroft, T., L.\ 2010, \apj, 710, 1578

Gregorio-Hetem, J., Montmerle, T., Rodrigues, C.~V., Marciotto, E., Preibisch, T., Zinnecker, H.\ 2009, \aap, 506, 711

\bibitem[Gunn et al.(1998)]{gunn98}
Gunn, J.~E., et al.\ 1998, \aj, 116, 3040 

\bibitem[Gunn et al.(2006)]{gunn06}
Gunn, J.~E., et al.\ 2006, \aj, 131, 2332 

\bibitem[Hennawi et al.(2006)]{hennawi06}
Hennawi, J.~F. et al.\ 2006, \aj, 131, 1

\bibitem[Hewett et al.(1998)]{hewett98}
Hewett, P.~C., Foltz, C.~B. \& Harding, M.~E., \ 2005, \aj, 115, 383

\bibitem[Hogg et al.(2001)]{hogg01}
Hogg, D.~W., Finkbeiner, D.~P., Schlegel, D.~J., \& 
Gunn, J.~E.\ 2001, \aj, 122, 2129 

\bibitem[Hopkins et al.(2005)]{hopkins05}
Hopkins, P.~F., Hernquist, L., Cox, T.~J., Di Matteo, T., Martini, P., Robertson, B., Springel, V., \apj, 630, 705  

\bibitem[Hopkins et al.(2006)]{hopkins06}
Hopkins, P.~F., Hernquist, L., Cox, T.~J., Di Matteo, T., Robertson, B., Springel, V., \apjs, 163, 1  

\bibitem[Hopkins et al.(2007)]{hopkins07}
Hopkins, P.~F., Bundy, K., Hernquist, L., Ellis, R.~S., \apj, 659, 976

\bibitem[Hopkins et al.(2008)]{hopkins08}
Hopkins, K., Hernquist, Cox, T.~J., Kere\v{s}, D., \apjs, 175, 356  

\bibitem[Ichikawa et al.(2006)]{ichikawa06}
Ichikawa, T., et al.\ 2006, \procspie, 6269, 38

\bibitem[Impey et al.(2002)]{impey02}
Impey, C.~D., Petry, C.~E., Foltz, C.~B., Hewett, P.~C.,\& Chaffee, F.~H. \ 2002, \apj, 574, 623 

\bibitem[Inada et al.(2003)]{inada03}
Inada, N., et al.\ 2003, \nat, 426, 810

\bibitem[Inada et al.(2005)]{inada05}
Inada, N., et al.\ 2005, \pasj, 57, L7

\bibitem[Inada et al.(2006)]{inada06}
Inada, N., et al.\ 2006, \apjl, 653, L97

\bibitem[Inada et al.(2008)]{inada08}
Inada, N., et al.\ 2008, \aj, 135, 496 

\bibitem[Inada et al.(2010)]{inada10}
Inada, N., et al.\ 2010, \aj, 140, 403 

\bibitem[Inada et al.(2012)]{inada12}
Inada, N., et al.\ 2012, \aj, 143, 119

\bibitem[Ivezi{\'c} et al.(2004a)]{ivezic04a}
Ivezi{\'c}, {\v Z}., et al.\ 2004a, Astronomische Nachrichten, 325, 583

\bibitem[Ivezi{\'c} et al.(2004b)]{ivezic04b}
Ivezi{\'c}, {\v Z}., Lupton, R.~H., Juric, M., et al. 2004b, in IAU Symp. 222, The Interplay among Black Holes, Stars and ISM in Galactic Nuclei, ed. Th. Storchi Bergmann, L.C. Ho \& H.R. Schmitt (Cambridge: Cambridge Univ. Press), 525

\bibitem[Iye et al.(2004)]{iye04}
Iye, M., et al.\ 2004, PASJ, 56, 381

\bibitem[Kayo \& Oguri(2012)]{kayo12}
Kayo, I., Oguri, M.\ 2012, \mnras, 424, 1363 

\bibitem[Kayser et al.(1986)]{kayser86}
Kayser, R., Refsdal, S., \& Stabell, R.\ 1986, \aap, 166, 36 

\bibitem[Keeton et al.(2000)]{keeton00}
Keeton, C.~R., Christlein, D., \& Zabludoff, A.~I.\ 2000, \apj, 545, 129 

\bibitem[Keeton et al.(2003)]{keeton03}
Keeton, C.~R.\ 2003, \apj, 584, 664  

\bibitem[Kochanek et al.(1999)]{kochanek99}
Kochanek, C.~S., Falco, E.~E., \& Mu\~{n}oz, J.~A.\ 1999, \apj, 510, 590 

\bibitem[Koopmans(2000)]{koopmans00}
Koopmans, L.~V.~E., at al. \aap, 2000, 361, 815

\bibitem[Landolt(1992)]{landolt92}
Landolt, A.~U.\ 1992, \aj, 104, 340

\bibitem[Leggett et al.(2006)]{leggett06}
Leggett, S.~K., et al.\ 2006, \mnras, 373, 781 

\bibitem[Lupton et al.(2004)]{lupton04}
Lupton, R., Blanton, M.~R., Fekete, G., Hogg, D.~W., O'Mullane, W., Szalay, A., Wherry, N.\ 2004, \pasp, 116, 133 

\bibitem[McLure \& Dunlop(2004)]{mclure04}
McLure, R.~J., Dunlop, J., S.\ 2004, \mnras, 352, 1390

\bibitem[More et al.(2012)]{more12}
More, A., Cabanac, R., More, S., Alard, C., Limousin, M, Kneib, J.~P., Gavazzi, R., Motta, V. \ 2012, \apj, 749, 38

\bibitem[Morgan et al.(2008)]{morgan08}
Morgan, C.~W., Eyler, M.~E., Kochanek C.~S., Morgan, N.~D., Falco, E.~E., Vuissoz, C., Courbin, F., Meylan, G.\ 1992, \apj, 676, 80

\bibitem[Mortlock et al.(1999)]{mortlock99}
Mortlock, D.~J., Webster, R.~L., \& Francis, P.~J.\ 1999, \mnras, 309, 836

\bibitem[Navarro et al.(1996)]{navarro96}
Navarro, J.~F., Frenk, C., S., White, S.~D.~M.\ 1996, \apj, 462, 563

\bibitem[Ofek et al.(2007)]{ofek07}
Ofek, E.~O., Oguri, M., Jackson, N., Inada, N., Kayo, I.\ 2007, \mnras, 382, 412

\bibitem[Oguri et al.(2004)]{oguri04}
Oguri, M., et al. \ 2004, \apj, 605, 78

\bibitem[Oguri et al.(2005)]{oguri05}
Oguri, M., et al.\ 2005, \apj, 622, 106 

\bibitem[Oguri et al.(2006)]{oguri06-1}
Oguri, M., et al.\ 2006, \aj, 132, 999 

\bibitem[Oguri(2006)]{oguri06-2}
Oguri, M. \ 2006, \mnras, 367, 1241

\bibitem[Oguri et al.(2008a)]{oguri08-1}
Oguri, M., et al.\ 2008a, \aj, 135, 512  

\bibitem[Oguri et al.(2008b)]{oguri08-2}
Oguri, M., et al.\ 2008b, \apj, 676, L1  

\bibitem[Oguri(2010)]{oguri10}
Oguri, M.\ 2010, \pasj, 62, 1017 

\bibitem[Oguri et al.(2012a)]{oguri12-1}
Oguri, M., et al.\ 2012a, \mnras, 420, 3213 

\bibitem[Oguri et al.(2012b)]{oguri12-2}
Oguri, M., et al.\ 2012b, \aj, 143, 120

\bibitem[Padmanabhan et al.(2004)]{padmanabhan04}
Padmanabhan, N., et al.\ 2004, New Astronomy, 9, 329 

\bibitem[Padmanabhan et al.(2008)]{padmanabhan08}
Padmanabhan, N., et al.\ 2008, \apj, 674, 1217 

\bibitem[Peng et al.(2002)]{peng02}
Peng, C.~Y., Ho, L.~C., Impey, C.~D., \& Rix, H.-W.\ 2002, \aj, 124, 266 

\bibitem[Peterson(1997)]{peterson97}
Peterson, B.~M.,\ 1997, An introduction to active galactic nuclei, Cambridge University Press 

\bibitem[Pier et al.(2003)]{pier03}
Pier, J.~R., Munn, J.~A., Hindsley, R.~B., Hennessy, G.~S., 
Kent, S.~M., Lupton, R.~H., 
\& Ivezi{\'c}, {\v Z}.\ 2003, \aj, 125, 1559 

\bibitem[Richards et al.(2002)]{richards02}
Richards, G.~T., et al.\ 2002, \aj, 123, 2945 

\bibitem[Richards et al.(2004)]{richards04}
Richards, G.~T., et al.\ 2004, \aj, 610, 679 

\bibitem[Rusin et al.(2003)]{rusin03}
Rusin, D., et al. \ 2003, \apj, 587, 143

\bibitem[Rusin \& Kochanek(2005)]{rusin05}
Rusin, D., Kochanek, C.~S.\ 2005, \apj, 623, 666


\bibitem[Schlegel et al.(1998)]{schlegel98}
Schlegel, D.~J., Finkbeiner, D.~P., \& Davis, M.\ 1998, \apj, 500, 525 


\bibitem[Schneider et al.(2006)]{schneider06}
Schneider, P., Kochanek, C., \& Wambsganss, J.\ 2006, Gravitational Lensing: Strong, Weak and Micro, Saas-Fee Advanced Course 33, Springer

\bibitem[Schneider et al.(2010)]{schneider10}
Schneider, D.~P., et al.\ 2010, \aj, 139, 2360 



\bibitem[Smith et al.(2002)]{smith02}
Smith, J.~A., et al.\ 2002, \aj, 123, 2121

\bibitem[Stoughton et al.(2002)]{stoughton02}
Stoughton, C., et al.\ 2002, \aj, 123, 485

\bibitem[Suzuki et al.(2008)]{suzuki08}
Suzuki, R., et al.\ 2008, \pasj, 60, 1347

\bibitem[Treu et al.(2006)]{treu06}
Treu, T, Koopmans, L.~V., Bolton, A.,~S., Burles, S., Moustakas, \& L.~A.\ 1984, \apj, 640, 662 

\bibitem[Tucker et al.(2006)]{tucker06}
Tucker, D.~L., et al.\ 2006, Astronomische Nachrichten, 327, 821 

\bibitem[Tully \& Fisher(1977)]{tully77}
Tully, R.~B., Ostriker, J.~P., \& Fisher, J.~R.\ 1977, \aap, 54, 661

\bibitem[Turner et al.(1984)]{turner84}
Turner, E.~L., Ostriker, J.~P., \& Gott, J.~R.\ 1984, \apj, 284, 1 

\bibitem[Vanden Berk et al.(2004)]{vandenberk04}
Vanden Berk, A., et al.\ 2004, \apj, 601, 692

\bibitem[Vilenkin(1984)]{vilenkin84}
Vilenkin, A.\ 1984, \apj, 282, L51

\bibitem[Walsh et al.(1979)]{walsh79}
Walsh, D., Carswell, R.~F., \& Weymann, R.~J.\ 1979, \nat, 279, 381

\bibitem[Weedman et al.(1982)]{weedman82}
Weedman, D.~W., Weymann, R.~J., Green, R.~F., \& Heckman, T.~M.\ 1982, \apjl, 255, L5

\bibitem[Wright et al.(2010)]{wright10}
Wright, E.~L., et al.\ 2010, \aj, 140, 1868

\bibitem[Wucknitz et al.(2003)]{wucknitz03}
Wucknitz, O., Wisotzki, L., Lopez, S., \& Gregg, M.~D.\ 2004, \aap, 405, 445

\bibitem[Yonehara et al.(2008)]{yonehara08}
Yonehara, A., Hirashita, H., Richter, P.\ 2008, \aap, 478, 95

\bibitem[York et al.(2000)]{york00}
York, D.~G., et al.\ 2000, \aj, 120, 1579

\end{thebibliography}
\end{document}